\documentclass[aps,prd,twocolumn,noshowpacs,superscriptaddress,groupedaddress,nofootinbib,preprintnumbers]{revtex4}
\pdfoutput=1
\usepackage{amsmath}
\usepackage{amsfonts}
\usepackage{amssymb}
\usepackage{graphicx, rotating}
\usepackage{epstopdf}
\usepackage{epsfig}
\usepackage{empheq}
\usepackage{latexsym}
\usepackage{multirow}
\usepackage{color}
\usepackage{amsmath,amssymb}
\usepackage{slashed}
\usepackage[colorlinks=true,linkcolor=red,anchorcolor=black,citecolor=blue,filecolor=cyan,menucolor=red,runcolor=filecolor,urlcolor=blue,bookmarks=true,bookmarksnumbered=true]{hyperref}%
\definecolor{rossos}{cmyk}{0,1,1,0.55}
\definecolor{bluscuro}{rgb}{0.15, 0.2, .85}
\definecolor{bluchiaro}{cmyk}{1,.3,0.,0.1}

\definecolor{rossos}{cmyk}{0,1,1,0.55}
\definecolor{bluscuro}{rgb}{0.15, 0.2, .85}
\definecolor{bluchiaro}{cmyk}{1,.3,0.,0.1}
\DeclareMathOperator{\diag}{diag}
\DeclareMathOperator{\GeV}{GeV}
\DeclareMathOperator{\TeV}{TeV}

\newcommand{\bc}{\begin{center}}
\newcommand{\ec}{\end{center}}
\newcommand{\bea}{\begin{eqnarray}}
\newcommand{\eea}{\end{eqnarray}}

\catcode`@=11
\def\marginnote#1{}

\newcount\hour
\newcount\minute
\newtoks\amorpm
\hour=\time\divide\hour by60
\minute=\time{\multiply\hour by60 \global\advance\minute by-\hour}
\edef\standardtime{{\ifnum\hour<12 \global\amorpm={am}%
\else\global\amorpm={pm}\advance\hour by-12 \fi
\ifnum\hour=0 \hour=12 \fi
\number\hour:\ifnum\minute<10 0\fi\number\minute\the\amorpm}}
\edef\militarytime{\number\hour:\ifnum\minute<10 0\fi\number\minute}
\def\draftlabel#1{{\@bsphack\if@filesw {\let\thepage\relax
\xdef\@gtempa{\write\@auxout{\string
\newlabel{#1}{{\@currentlabel}{\thepage}}}}}\@gtempa
\if@nobreak \ifvmode\nobreak\fi\fi\fi\@esphack}
\gdef\@eqnlabel{#1}}
\def\@eqnlabel{}
\def\@vacuum{}
\def\draftmarginnote#1{\marginpar{\raggedright\scriptsize\tt#1}}
\def\draft{\oddsidemargin 0.0truein
\def\@oddfoot{\sl ES, preliminary notes \hfil
\rm\thepage\hfil\sl\today\quad\militarytime}
\let\@evenfoot\@oddfoot \overfullrule 3pt
\let\label=\draftlabel
\let\marginnote=\draftmarginnote
\def\@eqnnum{(\theequation)\rlap{\kern\marginparsep\tt\@eqnlabel}%
\global\let\@eqnlabel\@vacuum} }
\catcode`@=12
\newcommand{\be}{\begin{equation}}
\newcommand{\ee}{\end{equation}}

\newcommand{\eq}[1]{Eq.~(\ref{#1})}
\def\l{\label}
\def\La{\mathcal{L}}

\def\({\left(}
\def\){\right)}
\def\<{\langle}
\def\>{\rangle}
\def\f{\frac}
\def\be{\begin{equation}}
\def\ee{\end{equation}}
\def\bry{\begin{array}}
\def\ery{\end{array}}
\def\bes{\begin{subequations}}
\def\ees{\end{subequations}}
\def\bit{\begin{itemize}}
\def\eit{\end{itemize}}
\def\ben{\begin{enumerate}}
\def\een{\end{enumerate}}
\def\dst{\displaystyle}

\def\de{\partial}
\def\demub{\de_{\mu}}

\def\ovl{\overline}
\def\Tr{\text{Tr}}

\def\psl{\slashed{p}}
\def\Dsl{\slashed{D}}

\definecolor{grey}{rgb}{0.6,0.6,0.6}
\definecolor{fuchsia}{rgb}{1,0,1}
\newcommand{\red}[1]{{\color{magenta}\color{red}#1\color{magenta}}}
\newcommand{\blue}[1]{{\color{blue}\color{blue}#1\color{blue}}}

\newcommand{\fuchsia}[1]{{\color{fuchsia}\color{fuchsia}#1\color{fuchsia}}}

\makeatother
\begin{document}

\preprint{CERN-PH-TH/2013-183}
\preprint{DFPD-2013/TH/14}
\preprint{LPN13-054}

\title{Higgs Couplings in Composite Models}
%

\author{Marc Montull}
\email{mmontull@ifae.es}
\address{IFAE, Universitat Aut\'onoma de Barcelona, 08193 Bellaterra, Barcelona, Spain}
\author{Francesco Riva}
\email{friva@ifae.es}
\address{Institut de Th\'eorie des Ph\'enom\`enes Physiques, EPFL,  CH--1015 Lausanne, Switzerland}
\author{Ennio Salvioni}
\email{ennio.salvioni@cern.ch}
\address{Theory Division, Physics Department, CERN, CH-1211 Geneva 23, Switzerland}
\address{Dipartimento di Fisica e Astronomia, Universit\`a di Padova, and \\ INFN Sezione di Padova, Via Marzolo 8, I-35131 Padova, Italy}
\author{Riccardo Torre}
\email{riccardo.torre@pd.infn.it}
\address{Dipartimento di Fisica e Astronomia, Universit\`a di Padova, and \\ INFN Sezione di Padova, Via Marzolo 8, I-35131 Padova, Italy}
\address{SISSA, Via Bonomea 265, I-34136 Trieste, Italy}
\begin{abstract}
We study Higgs couplings in the composite Higgs model based on the coset $SO(5)/SO(4)$. We show that the couplings to gluons and photons are insensitive to the elementary-composite mixings and thus not affected by light fermionic resonances. Moreover, at leading order in the mixings the Higgs couplings to tops and gluons, when normalized to the Standard Model (SM), are equal. These properties are shown to be direct consequences of the Goldstone symmetry and of the assumption of partial compositeness. In particular, they are independent of the details of the elementary-composite couplings and, under the further assumption of $CP$ invariance, they are also insensitive to derivative interactions of the Higgs with the composite resonances.
We support our conclusions with an explicit construction where the SM fermions are embedded in the $\mathbf{14}$ dimensional representation of $SO(5)$.
\end{abstract}

\maketitle

\section{Introduction}
\vspace{-2.5mm}
Models in which the Higgs boson arises as a composite resonance from a strongly coupled sector can provide a natural explanation of the small value of the electroweak (EW) scale. If in addition the Higgs is a pseudo-Nambu-Goldstone Boson (pNGB) associated to a spontaneously broken global symmetry, then the small value observed for its mass, compared to the mass of the other yet unobserved resonances, can be naturally explained. The minimal realization of this idea, known as Minimal Composite Higgs Model (MCHM), is based on the coset structure $SO(5)/SO(4)$ \cite{Agashe:2004ib,Contino:2006fd}. 

\vspace{-1mm}
The most promising signatures of these models are provided by the fermionic resonances, which are tightly connected to the EW scale because they are responsible for cutting off divergent contributions to the Higgs potential \cite{Matsedonskyi:2012ws,Redi:2012vq,Marzocca:2012tt,Pomarol:2012vn,Panico:2012vr,Pappadopulo:2013wt}. The bounds on heavy vector-like quarks from LHC direct searches are approaching the TeV \cite{CMS-PAS-B2G-12-012,CMS-PAS-B2G-12-015}. These resonances are also expected to modify the couplings of the Higgs to Standard Model (SM) particles and in particular the loop-induced couplings to gluons and photons. 
Information extracted from experimental analyses of Higgs couplings \cite{ATLAS-CONF-2013-034,CMS-PAS-HIG-13-005} can usefully complement the one coming from direct searches in constraining the natural parameter space of these models \cite{Falkowski:2013vg,Giardino:2013tu}.
Generically, the pNGB nature of the Higgs implies that the resonance contributions to the loop induced couplings, related to operators of the form $H^\dagger H F_{\mu\nu}F^{\mu\nu}$ that explicitly break the shift symmetry, are suppressed by powers of $g_{\text{SM}}/g_{\rho}$, where $g_{\rho}$ is a characteristic strong coupling \cite{Giudice:1024017,Low:2009gl}. The question then is whether or not this suppression disappears in the limit in which some of the resonances $\Psi$ are lighter than the others, $g_\Psi\ll g_\rho$. Surprisingly, this is not the case for a broad class of composite Higgs models, where light fermionic resonances do not contribute to the $hgg$ and $h\gamma\gamma$ couplings as the consequence of an exact cancellation between corrections to the $ht\bar{t}$ coupling and loops of resonances \cite{Falkowski:2007kl,Low:2010hn,2011arXiv1110.5646A}. In a two-site realization of the MCHM, this cancellation was shown to hold when only one Left-Right (LR) $SO(4)$ invariant is present \cite{2011arXiv1110.5646A}. In this paper we show that in the MCHM the cancellation, and therefore the insensitivity to light resonances, follows automatically from the pNGB nature of the Higgs and the assumption of partial compositeness, while it is independent of the number of LR $SO(4)$ invariants and of the particular realization of the elementary-composite couplings. Moreover, we find that under the further assumption of $CP$ conservation, derivative interactions of the Higgs to the resonances do not contribute to the $hgg$ and $h\gamma\gamma$ couplings. We show that the $ht\bar{t}$ and $hgg$ couplings are both fixed uniquely by the top mass, and coincide for small elementary/composite mixings. We also discuss how, in models where more than one LR $SO(4)$ invariant is present, these couplings are sensitive to the details of the UV physics \cite{2011arXiv1110.5646A}, even in the case where all resonances are heavy and possibly out of the direct reach of the LHC. 

\vspace{-1mm}
The paper is organized as follows. In Section \ref{sec:general} we present the general approach to the Higgs couplings in composite models based on the Callan-Coleman-Wess-Zumino (CCWZ) construction \cite{Coleman:1969p1798,Callan:1969p1799}. In Section \ref{sec:2sites} we describe an explicit realization based on a two-site version of the MCHM$_{14}$ \cite{Pomarol:2012vn,Panico:2012vr,Pappadopulo:2013wt} where the general features discussed in Section \ref{sec:general} are exemplified. We also comment on an alternative approach based on the Weinberg Sum Rules (WSR) \cite{Marzocca:2012tt,Pomarol:2012vn}. Finally, in Section \ref{sec:conclusion} we draw our conclusions. Appendices \ref{app:notations} and \ref{app:fit} contain a summary of our notations and details on our fit to Higgs data, respectively.

\vspace{-3mm}
\section{General Composite Higgs Models} \label{sec:general}
The scalar sector of minimal pNGB Higgs models, based on the $SO(5)/SO(4)$ coset structure\footnote{An extra unbroken $U(1)_{X}$ is always understood in the coset structure, in order to reproduce the correct hypercharge of the SM fermions. Our normalization is such that $Y=T_{R}^{3}+X$.}, is described by the Goldstone matrix
\vspace{-1mm}\be\label{GB}
U(\Pi)=\exp\left(i\frac{\sqrt{2} \Pi^{i} T^{i}}{f}\right)\,.
\ee
where $T^{i}$ are the broken generators, $\Pi^{i}$ the Goldstone bosons and $f$ the corresponding decay constant (see Appendix~\ref{app:notations} for the notation).

We assume that the SM fermions obtain their masses through {\it partial compositeness} \cite{1991NuPhB.365..259K}, by mixing with operators of the strong sector $\mathcal{O}^{\,I,...,J}$, with capital letters $I,J$ denoting $SO(5)$ indices. This mixing is conveniently described by formally promoting the elementary fermions to full representations of the $SO(5)$ group, the \emph{embeddings}. The embeddings for the $SU(2)_{L}$ doublet $q_L$ and for the singlet $q_R$ are denoted by $\mathcal{Q}_{L}^{I,\cdots,J}\,$, $\mathcal{Q}_{R}^{I,\cdots,J}$, respectively.
%
Then the Lagrangian for partial compositeness takes the form
\vspace{-2mm}
\begin{equation}\label{OpsMix}
\mathcal{Q}_{L,R}^{I\cdots J}\mathcal{O}_{I\cdots J}\,.
\end{equation}
At low energy, in the broken phase, this implies mixing terms between elementary fermions and resonances of the strong sector $\Psi_r$, which, up to small splittings proportional to the EW symmetry breaking {\scshape vev}, can be taken as full multiplets $\mathbf r$ of  the unbroken $SO(4)$. A convenient way to write these low-energy interactions, while keeping track of the underlying $SO(5)$ symmetry, exploits the transformation properties of the Goldstone matrix  $U\to g\, U \, \hat{h}(g ,\Pi)^{-1}$ with $g \in SO(5)$, $\hat{h} \in SO(4)$ \cite{Coleman:1969p1798,Callan:1969p1799}. In fact, the Goldstone matrix $U$ can be used to `convert' irreducible representations of $SO(5)$ into reducible representations of $SO(4)$. Then one can write interactions between the embeddings, transforming under $SO(5)$, and the resonances in representations of $SO(4)$:
\begin{equation}\label{generalMixing}
\hspace{-2mm}\bry{lll}
\dst \mathcal{L}_{\mathrm{mix}}&=&\dst \Big( F^{L}_r \overline{\mathcal{Q}}_{L}^{I\cdots J} U_{Ii}...U_{Jj}\Psi_r^{i\cdots j}+\cdots\vspace{1mm}\\
&&\dst \hspace{-2mm}+F^{L}_1 \overline{\mathcal{Q}}_{L}^{I\cdots J} U_{I5}...U_{J5}\Psi_1+\textrm{h.c} \Big) +\left(L\to R\right)\,\,,
\ery
\end{equation}
where the dots stand for couplings with resonances in other $SO(4)$ representations. The simplest example of \eq{generalMixing} is the MCHM$_{5}$ \cite{Contino:2006fd}: in this case \eq{OpsMix} implies the existence of resonances in a $\bf 5={\bf 4} \oplus {\bf 1}$ of $SO(4)\,$, and \eq{generalMixing} reads $\,F^L_4  \overline{\mathcal{Q}}_{L}^{\,I}U_{Ii}\Psi_4^{i}+F^L_1 \overline{\mathcal{Q}}_{L}^{\,I} U_{I5}\Psi_1 + \mathrm{h.c.} + \left(L\to R\right)$.

We further assume that some of the fermionic resonances are lighter than the typical scale of the other resonances. This assumption is motivated by the tension between the necessary scale of bosonic resonances which, to account for the smallness of the $S$ parameter, are expected to be in the multi-TeV range, and the need for light fermionic resonances as necessary to reproduce the smallness of the observed Higgs mass \cite{Matsedonskyi:2012ws,Marzocca:2012tt,Pomarol:2012vn,Redi:2012vq,Panico:2012vr,Pappadopulo:2013wt}. In this limit we can keep some of the resonances in our effective description, while decoupling the heavy ones. The Lagrangian describing this setup contains, in addition to \eq{generalMixing}, a part describing the strong sector alone, which can be written, again, with the $SO(5)$ symmetry non-linearly realized \cite{Coleman:1969p1798,Callan:1969p1799},
\begin{equation}\label{dterms}
\hspace{-3mm}\bry{lll}
\dst \mathcal{L}_{\mathrm{strong}}&=&\dst (\mathrm{kin.\,term\,for}\,\Psi_r)-M_r\overline{\Psi}_r\Psi_r+\cdots\vspace{2mm}\\
&&\dst \hspace{-9.5mm}+i c_{L}\,\overline{\Psi}_{rL}^{\,i\cdots jk}\gamma^\mu d^k_\mu\Psi_{r^{\prime} L}^{i\cdots j}+\cdots+\textrm{h.c.} + (L\to R),
\ery
\end{equation}
where at leading order in the chiral expansion and in the unitary gauge $d_{\mu}^{\,k}\simeq (\sqrt{2}/f) \demub h\, \delta^{k4}$ (see Appendix \ref{app:notations} for details) and the dots stand for different representations. We have denoted by $\mathbf{r}$ and $\mathbf{r^\prime}$ two representations that differ by one $SO(4)$ index, in order to allow for the first term in parentheses.
In the example of the MCHM$_{5}$, the second line of \eq{dterms} reads $i c_{L}\,\overline{\Psi}_{4L}^{\,k}\gamma^\mu d^k_\mu\Psi_{1L}+\mathrm{h.c.}+\left(L\to R\right)$.

Notice that a number $n_{r}>1$ of copies of each $SO(4)$ multiplet could be present in the low-energy theory. In this case, mass mixing terms between the $\Psi_{r}^{(i)}\,$, \mbox{($i=1,\ldots, n_{r}$)} are allowed by the global symmetry. However, these mass mixings can always be eliminated with a suitable field redefinition, so the masses in the strong sector can be taken diagonal without loss of generality. 
In Eq.~\eqref{dterms} we have neglected higher derivative interactions: beside being suppressed by the strong sector scale, these interactions do not affect the couplings of a single Higgs with a pair of gauge bosons. 

The terms in the second line of \eq{dterms} couple $h$ with two resonances and can potentially give sizeable corrections to the $hgg$ coupling. However, as we now show, if a further assumption is made on the theory, namely $CP$ conservation, then the contribution to the $hgg$ coupling of these operators exactly vanishes. Indeed, if the coefficient $c_{L}$ is real then the Higgs derivative interactions contained in \eq{dterms} can be written as
\begin{equation} \label{eq: deriv terms}
i \sqrt{2} c_{L}\frac{\partial_{\mu}h}{f}\, \left(\overline{\Psi}_{rL}^{\,i\ldots j4}\gamma^{\mu}\Psi_{r^{\prime}L}^{i\ldots j} - \overline{\Psi}_{r^{\prime}L}^{\,i\ldots j}\gamma^{\mu} \Psi_{rL}^{\,i\ldots j4}\right)\,.
\end{equation}
The interactions in \eq{eq: deriv terms} are manifestly \emph{antisymmetric}\footnote{In the argument that follows we neglect, without loss of generality, the phases that appear in the definitions of the composite multiplets (see for example Eq.~\eqref{composite1-4}).} in the fermion fields (notice that because $d_{\mu}$ transforms as a $\mathbf{4}$ of $SO(4)$, Eq.~\eqref{dterms} does not generate Higgs derivative interactions that are bilinear in the same fermion field). Equation \eqref{eq: deriv terms} is written in the gauge eigenstate basis for fermions. Now, if the parameters that appear in the fermion mass matrix, namely the masses $M_{r}$ and the linear mixings $F_{r}^{L,R}$, are also real, then the unitary transformations that diagonalize the mass Lagrangian are orthogonal, and the Higgs derivative couplings are antisymmetric in the mass eigenstate basis as well. Because the gluon only has diagonal couplings, however, vertices involving the Higgs and two distinct fermions do not contribute to the triangle one-loop diagrams for $hgg$. Thus we conclude that under the hypothesis that the Lagrangian preserves a $CP$ symmetry, the operators in Eq.~\eqref{dterms} do not contribute to single Higgs production.\footnote{The situation is different in the double Higgs production process, $gg\to hh\,$, because couplings of the Higgs to two distinct fermions can enter in box diagrams \cite{Gillioz:2012vn}.} Alternatively, if $CP$ is not preserved a contribution to the $hgg$ coupling generically arises, proportional to $G_{\mu\nu}^{A}\widetilde{G}^{\mu\nu\,A}\,$. A completely analogous argument holds for the $h\gamma\gamma$ coupling. As an example, let us consider the top sector of the MCHM$_{5}$. Assuming that all the parameters in Eq.~\eqref{dterms} are real but allowing for complex linear mixings in Eq.~\eqref{generalMixing}, the Higgs derivative interactions that contribute to the $hgg$ coupling, obtained transforming Eq.~\eqref{eq: deriv terms} into the mass eigenstate basis, read
\be
c_{L}\f{\demub h}{f}\sum_{a\,=\,1}^{4}k_{L}^{a}\,\ovl{\psi}_{L}^{\,a}\gamma^{\mu} \psi_{L}^{a}\,,
\ee
where $\psi^{a}$ are the mass eigenstate fermions (that is, the physical top quark and its partners) and the leading contributions to the coefficients $k_{L}^{a}$ scale as
\be
\f{\text{Im}(F_{4}^{L,R\,\ast}F_{1}^{L,R})}{M_{1}M_{4}}\sqrt{\xi}\,.
\ee
For simplicity, since in this paper we focus on the top sector, from here on we assume that all the parameters in the Lagrangian can be made real by redefining the fermion fields. Under this assumption, therefore, we conclude that the terms in Eq.~\eqref{dterms} have no impact on the couplings between the Higgs and massless gauge bosons. Of course, when all the SM fermions are included, a source of $CP$ violation must be present to reproduce the Cabibbo-Kobayashi-Maskawa phase. In this case the strong constraints on $CP$-odd observables from flavor physics \mbox{\cite{Csaki:2008dj,Redi:2011wr,KerenZur:2012td,Barbieri:2012wia}} should be taken into account.

From the above discussion we conclude that the $hgg$ coupling is determined by Higgs interactions at zero momentum. In this limit, the coupling of the Higgs to gluons mediated by loops of a particle with mass $M\gg m_{h}$ can be derived from the contribution of the heavy particle to the QCD $\beta$ function, by means of the Higgs low-energy theorem \cite{1976NuPhB.106..292E,Shifman:CQH_Aq6g}. Therefore, neglecting the contribution of the light SM fermions we simply have (for each SM particle $x$ we define $c_{x}\equiv g_{hxx}/g_{hxx}^{\text{SM}}$)
\begin{equation} \label{eq: hgg general}
\hspace{-1mm}c_{g}\hspace{-1mm}= \hspace{-1mm}\frac{v}{2}\bigg[\frac{\partial}{\partial h}\log \det \mathcal{M}_{t}^{\dagger}\mathcal{M}_{t}(h) + \sum_{i} \frac{\partial}{\partial h} \log M_{f,i}^{2}(h)\bigg]_{\left\langle h \right\rangle}\,,
\end{equation}
where $\mathcal{M}_{t}(h)$ is the mass matrix in the top sector, and we have also included the contribution from the partners of the light SM fermions, with squared masses $M_{f,i}^{2}(h)$. Fermions with `exotic' electric charges (such as for example $Q_{el}=5/3$ or $8/3$, which are present in composite Higgs models) do not contribute to the $hgg$ coupling, because since they do not mix with the elementary fermions, they do not feel any explicit breaking of the $SO(5)$ symmetry; as a consequence, loops involving only the exotic states cannot generate any effects that break the shift symmetry, including a $hG_{\mu\nu}^{A}G^{\mu\nu\,A}$ coupling.

Let us focus on the contribution arising from the top sector.\footnote{The partners of a light SM fermion $f$ give a contribution to $c_{g}-1$ that scales like $\sim \epsilon_{f\,L,R}^2 \,\xi\,$, where $\epsilon_{f\,L,R}$ measure the degree of compositeness of $f_{L,R}$, and is thus competitive with the $\sim \xi$ contribution of the top sector only in the limit of full compositeness for one of the chiralities of $f$ \cite{2011arXiv1110.5646A,Delaunay:2013us}. Therefore, for a generic point in parameter space, the contribution of the partners of the light SM fermions is expected to be subleading. See for example Eq.~\eqref{eq:cg full} in the following.} Assuming the presence of $n$ top partners in the theory, the mass Lagrangian in the top sector can be written in full generality as
\begin{equation} \label{eq: pc mass lagr}
\vspace{-3mm}
-\(\bry{c|ccc} \bar{t}_{L} &\quad& \hspace{-3mm}\overline{\mathbf{C}}_{L}\hspace{-3mm}&\quad\ery\) \mathcal{M}_{t}(h)  \(\bry{cccc} t_{R} \\\hline \vspace{-3mm}\\ \mathbf{C}_{R}\\\vspace{-3mm}\quad \ery\) + \mathrm{h.c.}\,
\vspace{-2mm}
\end{equation}
with
\begin{equation} \label{eq: pc mass lagr2}
\mathcal{M}_{t}(h)=\(\bry{c|ccc} 0&\quad&\hspace{-3mm}\mathbf{F}_{L}^{T}(h)\hspace{-3mm}&\quad\\\hline \quad&\quad&\quad&\quad\vspace{-3mm}\\ \mathbf{F}_{R}(h)&\quad&\mathbf{M}_{c}&\quad\\
\ery\)\,,
\end{equation}
where $\mathbf{C}$ is a $n$-dimensional vector collecting all the top partners, and $\mathbf{F}_{L,R}(h)$ are $n$-dimensional vectors containing the elementary-composite mixing terms. Since, by assumption, the only breaking of the global symmetry under which the Higgs shifts is contained in the mixings with elementary states, the strong sector alone can only generate derivative interactions of $h$ and the $n\times n$ mass matrix in the composite sector $ \mathbf{M}_{c}$ is independent of the Higgs field. Thus the structure in Eq.~\eqref{eq: pc mass lagr} follows from the assumption of partial compositeness. From the properties of block matrices we find
\begin{equation} \label{eq: hgg}
\det \mathcal{M}_{t}(h)= m_{t}^{0}(h) \times \det \mathbf{M}_{c}\,,
\ee
which implies that the contribution to the $hgg$ coupling from the top sector is
\be\l{equation10}
\dst c_{g}^{(t)} = v \bigg[\frac{\partial}{\partial h}\log m_{t}^{0}(h)\bigg]_{\left\langle h \right\rangle}\,.
\end{equation}
Here $m_{t}^{0}(h)=-\mathbf{F}_{L}^{T}(h) \mathbf{M}_{c}^{-1}\mathbf{F}_{R}(h)$ is the top mass at quadratic order in the mixings $F_{r}^{L,R}\,$, which can be readily obtained from Eq.~\eqref{eq: pc mass lagr} by integrating out the composite states:
\begin{equation} \label{eq: eff lagr top}
\mathcal{L}_{\mathrm{eff}}= - m_{t}^{0}(h)\bar{t}_{L}t_{R}+\mathrm{h.c.} + i Z_{t_{L}}(h)\bar{t}_{L}\slashed{\partial}t_{L} + i Z_{t_{R}}(h)\bar{t}_{R}\slashed{\partial}t_{R}\,,
\end{equation}
where we have also included the renormalizations to the wavefunctions of $t_{L,R}\,$. From Eq.~\eqref{eq: eff lagr top} we derive the coupling of the Higgs to the top
\begin{equation} \label{eq: htt}
\bry{l}
\dst c_{t}= v \bigg[\frac{\partial}{\partial h}\log m_{t}(h)\bigg]_{\left\langle h \right\rangle}\,,\quad m_{t}(h)=\frac{m_{t}^{0}(h)}{\sqrt{Z_{t_{L}}(h)Z_{t_{R}}(h)}}\,.
\ery
\end{equation}
In the limit where the breaking of the global symmetry is small, $\epsilon_{L,R}^{2} \sim (F_{r}^{L,R}/M_{\Psi})^{2} \ll 1\,$ (where $M_{\Psi}$ generically denotes the composite masses), one finds \mbox{$Z_{t_{L,R}}(h)\sim 1 + \epsilon_{L,R}^{2}f_{L,R}(h)\,$}, where $f_{L,R}(h)$ are periodic functions of $h\,$, and we neglected terms of higher order in $\epsilon^{2}_{L,R}\,$. Thus the top contribution to the $hgg$ amplitude, Eq.~\eqref{equation10}, is tightly correlated with the $ht\bar{t}$ coupling in Eq.~\eqref{eq: htt}, and the two can differ sizably only if $\epsilon_{L,R}\,\sim1\,$, that is if one of the chiralities of the top is mostly composite. 
Furthermore and importantly, while the $ht\bar{t}$ coupling receives corrections at all orders in $\epsilon^{2}_{L,R}\,$, from Eq.~\eqref{equation10} we read that the $hgg$ coupling is formally of zeroth-order in this expansion. This implies that the terms of higher order in $\epsilon_{L,R}^{2}$ in the top loop contribution to the $hgg$ amplitude are exactly canceled by those coming from loops of resonances. This cancellation was found to take place in several models where the Higgs is a pseudo-Goldstone boson \cite{Falkowski:2007kl,Low:2010hn,2011arXiv1110.5646A}, and implies that the presence of light fermionic resonances would not affect the production rate of the Higgs via gluon fusion, nor its decay width into photons. Our analysis shows that, in the context of pNGB Higgs models, this result follows automatically from the assumption of partial compositeness, and is not dependent on the choice of the embedding for the elementary fermions nor on the specific realization of the model. Indeed, our analysis was performed by applying the general CCWZ approach.

Let us now inspect more closely the structure of the $hgg$ coupling. According to Eq.~\eqref{equation10}, its expression is determined by the LR $SO(4)$ invariants that can be built out of the embeddings $\mathcal{Q}_{L,R}$ and that contribute to the top mass $m_{t}^{0}\,$. The latter has the form
\begin{equation} \label{mt0 general}
m_{t}^{0}(h)=\sum_{n=1}^{N}\left( \sum_{r} c_{r}^{(n)} y_{r} \right)\times I_{LR}^{(n)}\left(\frac{h}{f}\right)\,,
\end{equation}
where $I_{LR}^{(n)}$ indicates the $N\geq 1$ $SO(4)$ invariants. The coefficient of each invariant is given by a linear combination of the quantities
\vspace{-2mm}
\begin{equation}\label{yr}
y_{r}\equiv \sum_{i=1}^{n_{r}} \frac{F_{r(i)}^{L}F_{r(i)}^{R}}{M_{r(i)}}\,,\vspace{-2mm}
\end{equation}
with coefficients $c_{r}^{(n)}$. This was expected, since $y_{r}$ is simply the leading contribution of the $\mathbf{r}$-plets to the top Yukawa coupling. From Eq.~\eqref{mt0 general} we readily obtain
\begin{equation} \label{eq: hgg pc}
c_{g}^{(t)} = 1 - \Delta_{g}^{(t)}(y_{r}/y_{r'})\,\xi + \mathcal{O}(\xi^{2}) \,,
\end{equation}
where $\Delta_{g}^{(t)}$ is a function with values of $\mathcal{O}(1)$ and $y_{r}/y_{r'}$ schematically denotes all the different ratios of $y_{r}$ that can be built in the chosen model. While this is indeed the most general form of the $hgg$ coupling, its expression further simplifies if only one LR invariant can be built out of the embeddings $\mathcal{Q}_{L,R}$, \emph{i.e.} if $N=1\,$ in Eq.~\eqref{mt0 general}. In this case, when taking $\partial \log m_{t}^{0}/\partial h$ in Eq.~\eqref{equation10}, the dependence on the $y_{r}$ drops and the $hgg$ coupling turns out to be a simple `trigonometric' rescaling of the SM expression. In other words, if $\,N=1\,$ then $\,\Delta_{g}^{(t)}\, =\, \mathrm{constant}\,$ in Eq.~\eqref{eq: hgg pc}. This was already noticed in Ref.~\cite{2011arXiv1110.5646A}, where a two-site setup was considered. For example, in the popular MCHM$_{5}$ and MCHM$_{10}\,$ \cite{Contino:2006fd} there is only one LR invariant:\footnote{Naively, in each of the products $\mathbf{5}_{L}\times \mathbf{5}_{R}$ and $\mathbf{10}_{L}\times \mathbf{10}_{R}$ \emph{two} $SO(4)$ invariants appear. However, whenever $q_{L}$ and $t_{R}$ are embedded in the same $SO(5)$ representation $\mathbf{r}$, one invariant does not depend on the Higgs and can be written as $\overline{\mathcal{Q}}_{L}\mathcal{Q}_{R}$, which vanishes when the embeddings are set to their physical values. Thus the number of invariants is lowered by one unit \cite{Mrazek:2011iu}.}
\begin{align} \nonumber
\mathbf{5}_{L,R}:&\,\quad U_{Ii}(\hat{Q}_{t_{L}}^{\dagger})_{I}(\hat{Q}_{t_{R}})_{J}U_{Ji} =\frac{1}{2\sqrt{2}}\,s_{2h}\,, \\
\mathbf{10}_{L,R}:&\,\quad U_{Ii} (\hat{Q}_{t_{L}}^{\dagger})_{IJ}(\hat{Q}_{t_{R}})_{JK}U_{Ki}=-\frac{1}{8}\, s_{2h}\,
\end{align}
where $s_{n h}\equiv\sin(n h/f)$ and we defined $\mathcal{Q}_{L}\equiv t_{L}\hat{Q}_{t_{L}}+b_{L}\hat{Q}_{b_{L}}$ and $\mathcal{Q}_{R}\equiv t_{R}\hat{Q}_{t_{R}}$. In both cases $I_{LR}^{(1)}=s_{2h}$, leading to
\begin{equation} \label{eq: hgg 5+5}
\mathbf{5}_{L,R}\,,\quad\mathbf{10}_{L,R}\,:\quad c_{g}^{(t)}= \frac{1-2\xi}{\sqrt{1-\xi}} \,\,\Rightarrow\,\, \Delta_{g}^{(t)} = \frac{3}{2}\,.
\end{equation}
On the other hand, two independent LR invariants are present for example in MCHM$_{14}\,$: 
\begin{equation}
\hspace{-1mm}\bry{l}
U_{I5}(\hat{Q}_{t_{L}}^{\dagger})_{IJ}U_{J5}U_{K5}(\hat{Q}_{t_{R}})_{KL} U_{L5}=\frac{1}{16\sqrt{5}}(-\,6\,s_{2h}-5 \,s_{4h}), \vspace{2mm}\\
U_{Ii}(\hat{Q}_{t_{L}}^{\dagger})_{IJ} U_{Jj}U_{Ki}(\hat{Q}_{t_{R}})_{KL} U_{Lj}\,=\frac{1}{16\sqrt{5}}(\,6\,s_{2h}-5 \,s_{4h})\,.\,\,
\ery
\end{equation}
It follows that the dependence on the $y_{r}$ does not drop out of the $hgg$ coupling, which takes the general form in Eq.~\eqref{eq: hgg pc}. By explicit computation we find  
\begin{equation}\label{eq18}
\mathbf{14}_{L,R}:\qquad \Delta_{g}^{(t)} = \frac{11}{2}\bigg(\frac{1-\frac{64}{55}\frac{y_{1}}{y_{4}}-\frac{6}{11}\frac{y_{9}}{y_{4}}}{1-\frac{8}{5}\frac{y_{1}}{y_{4}}}\bigg)\,.
\end{equation}
Contrarily to the models with only one invariant, where a single universal function of $\xi$ appears (see for example Eq.~\eqref{eq: hgg 5+5}), when $N>1$ a continuum of possible couplings to photons and gluons is allowed by the symmetry structure. Furthermore, while the `trigonometric' rescaling of models with a single invariant always suppresses the Higgs production rate, in models with more than one invariant $\Delta_{g}^{(t)}$ can take both signs depending on the values of the ratios $y_{r}/y_{r'}\,$, thus an enhancement of the rate is in principle also possible. However, notice that from Eq.~\eqref{eq18}, taking the limit where one $\mathbf{1}\,(\mathbf{4})$ is much lighter than all the other resonances,\footnote{When one $\mathbf{9}$ is much lighter than the other resonances one finds $m_{t}^{0}(h)\propto s_{h}^{3}c_{h}$ and therefore $c_{g}^{(t)}\simeq 3 - 5\xi/2\,$. Similarly, the $ht\bar{t}$ coupling is equal to $3$ times its SM value in the limit $\xi\to 0\,$. Thus we do not regard this possibility as viable.} we find $\Delta_{g}^{(t)}=4 \,\,(\Delta_{g}^{(t)}=11/2)\,$: in both cases the rate is actually strongly suppressed, suggesting that in most of the parameter space of the MCHM$_{14}$ we should expect $c_{g}^{(t)}<1\,$.\footnote{We expect the typical value of $\Delta_{g}^{(t)}$ to increase with the dimension of the $SO(5)$ representation. Therefore, for large enough representations negative values of $c_{g}^{(t)}$ might be possible.} This will be confirmed by the detailed analysis contained in Section~\ref{sec:2sites}. In Table~\ref{table:summaryCCWZ} we report the values of $\Delta_{g}^{(t)}$ for the lowest-dimensional embeddings compatible with the custodial symmetry that protects the $Z$-$b$-$\bar{b}$ coupling \cite{Agashe:2006at}. 
Notice that the results in the column corresponding to $\mathcal{Q}_{R}\sim \mathbf{1}$ hold even if the $t_{R}$ is assumed to be a fully composite chiral state, rather than an elementary field mixed with a strong sector operator. In fact, if $t_{R}$ is fully composite the structure of the mass matrix differs from that in Eq.~\eqref{eq: pc mass lagr2}, but Eq.~\eqref{equation10} still holds. Therefore, independently of whether $t_{R}$ is a partially or fully composite singlet of $SO(5)$, the $hgg$ coupling is determined by the $SO(4)$ invariants that can be built out of $\mathcal{Q}_{L}$ and the Goldstone matrix, and are linear in the former.  

\begin{table}[t]
		\centering
		\begin{tabular}{|c|c|c|c|c|} \hline
	 $\mathcal{Q}_{L}\, \backslash \,\mathcal{Q}_{R}$ & $\mathbf{1}$  &  $\mathbf{5}$ & $\mathbf{10}$ & $\mathbf{14}$ \\
	\hline & & & &
	\\[-0.4cm] $\mathbf{5}$  &  $1/2$ & $3/2$ & $1/2$ & $\dst \frac{5}{2}\,\frac{1-\frac{24}{25}\frac{y_{1}}{y_{4}}}{1-\frac{4}{5}\frac{y_{1}}{y_{4}}}$ \\[0.35cm]
	\hline
	$\mathbf{10}$ &  $\bigtimes$ & $1/2$ & $3/2$ & $3/2$ \\[0.05cm]
	\hline  & & & &
	\\[-0.4cm] $\mathbf{14}$ &  $3/2$ & $\dst \frac{9}{2}\,\frac{1-\frac{10}{9}\frac{y_{1}}{y_{4}}}{1-2\frac{y_{1}}{y_{4}}}$ & $3/2$ & $\dst \frac{11}{2}\,\frac{1-\frac{64}{55}\frac{y_{1}}{y_{4}}-\frac{6}{11}\frac{y_{9}}{y_{4}}}{1-\frac{8}{5}\frac{y_{1}}{y_{4}}}$ \\[0.35cm]
	\hline		
		\end{tabular}
		\caption{Summary table showing the value of $\Delta_{g}^{(t)}\,$, defined by Eq.~\eqref{eq: hgg pc}, for different choices of the embeddings of elementary fermions. The $y_{r}$ were defined in Eq.~\eqref{yr}. The points at which $\Delta_{g}^{(t)}$ formally diverges (for example, $y_{4}=8y_{1}/5$ for $\mathbf{14}_{L}+\mathbf{14}_{R}$) correspond to the nonviable situation where $m_{t}^{0}\propto s_{h}^{3}c_{h}$ and thus $c_{t}\to 3$ for $\xi\to 0\,$, \emph{i.e.} the SM top Yukawa is not recovered in the limit $\xi\to 0\,$. In the case $\mathcal{Q}_{L}\sim\mathbf{10}$, $\mathcal{Q}_{R}\sim \mathbf{1}$, there is no invariant that can generate the top mass.\vspace{-5mm}}
	\label{table:summaryCCWZ}
\end{table}
As first pointed out in Ref.~\cite{2011arXiv1110.5646A}, in models which feature more than one LR invariant, such as MCHM$_{14}$, the Higgs production rate is sensitive to the resonance spectrum, implying that the analysis of Higgs couplings can usefully complement the information coming from direct searches for heavy fermions. We note that because of the dependence on the ratios $y_{r}/y_{r'}$, the Higgs coupling to gluons is insensitive to the absolute scale of the resonances. Therefore one can envisage a finely-tuned scenario where all the top partners are relatively heavy and thus out of the direct reach of the LHC \cite{Panico:2012vr}, but the imprint they leave on Higgs rates still carries some information about UV physics. In this `split' version of the composite Higgs setup, the Higgs couplings would be the primary source of information about the strong sector.

It is important to observe that when the light generations are included in the theory, the presence of multiple $SO(4)$ invariants in the LR sector gives rise to Higgs-mediated FCNC at tree level \cite{Agashe:2009dg,Mrazek:2011iu}. These flavor-changing Higgs couplings are suppressed only by $\xi\,$, which is generically not enough to comply with bounds from flavor physics, such as Kaon mixing. This issue is relaxed if the underlying flavour structure realizes Minimal Flavor Violation (MFV) \cite{Redi:2011wr}. This would imply in particular a sizable degree of compositeness for one of the chiralities (either left or right) of all SM fermions, making the contribution of the partners of light quarks to the $hgg$ and $h\gamma\gamma$ couplings potentially sizable \cite{Delaunay:2013us}. 

\section{An explicit construction: MCHM$_{14}$}\l{sec:2sites}
\vspace{-2mm}
In this section we describe in detail one explicit model where the Higgs couplings to gluons and photons can take a continuum of values depending on the spectrum of resonances, as in Eq.~\eqref{eq: hgg pc}. As we discussed, this happens when the top mass arises from at least two independent $SO(4)$ invariants. Here we focus on the realization of the MCHM where both $q_{L}$ and $t_{R}$ are embedded into a $\mathbf{14}$ with $X$ charge equal to $2/3$:
\vspace{-1mm}
\be\label{spurions}
\bry{l}
\mathcal{Q}_{L}= \,\dst \f{1}{2} \begin{pmatrix} & & & & i b_{L} \\
& & & & b_{L} \\
& & & & i t_{L} \\
& & & & - t_{L} \\
i b_{L}& b_{L} & i t_{L} & - t_{L} &   \\ \end{pmatrix}\,,\vspace{2mm}\\
\dst \mathcal{Q}_{R}=\dst \f{1}{2\sqrt{5}} t_{R} \diag\,(-1,-1,-1,-1,4)\,.
\ery
\ee
%
We recall that $\mathbf{14}=\mathbf{9}\oplus\mathbf{4}\oplus\mathbf{1}$ under $SO(4)$. Including for simplicity only one copy of each composite multiplet $\Psi_{9,4,1}\,$, the Lagrangian for the top sector can be written in the form
\vspace{-1mm}
\begin{align}\l{2sitestop}
\La_{t} &\,=\, \dst i\ovl{q}_{L}\Dsl q_{L}+i\ovl{t}_{R}\Dsl t_{R} \,+ i \overline{\Psi}_{1}\Dsl \Psi_{1} \nonumber \\ 
&\hspace{-2mm}+i \overline{\Psi}_{4}(\Dsl + i\slashed{e}) \Psi_{4} +\,i \mathrm{Tr}[\overline{\Psi}_{9}(\Dsl \Psi_{9} + i[\slashed{e},\Psi_{9}])] \nonumber\\
&\hspace{-2mm}- M_{1}\,\ovl{\Psi}_{1}\Psi_{1} - M_{4}\,\ovl{\Psi}_{4}\Psi_{4} - M_{9} \Tr[\ovl{\Psi}_{9} \Psi_{9}] \nonumber\\
&\hspace{-2mm}+\Big(\, F^{L}_9\Tr[(U^{T}\,\overline{\mathcal{Q}}_{L}U)\Psi_{9R}]+ F^{R}_9\Tr[\overline{\Psi}_{9L} (U^{T}\mathcal{Q}_{R}U)] \nonumber\\
&\hspace{-2mm}+ \sqrt{2}F_{4}^{L}(U^{T}\overline{\mathcal{Q}}_{L}U)_{5i} (\Psi_{4R})_{i} + \sqrt{2}F_{4}^{R}(\overline{\Psi}_{4L})_{i}(U^{T}\mathcal{Q}_{R}U)_{i5} \nonumber\\
&\hspace{-2mm}+ \frac{\sqrt{5}}{2}F_{1}^{L}(U^{T}\,\overline{\mathcal{Q}}_{L}U)_{55}\,\Psi_{1R}+ \frac{\sqrt{5}}{2} F_{1}^{R}\, \overline{\Psi}_{1L} \,(U^{T}\mathcal{Q}_{R}U)_{55}\, \nonumber \\&\,+\, \mathrm{h.c.}\Big) \,,
\end{align}
where $D^{\mu}\Psi_{r}\equiv (\partial^{\mu}-i g^{\prime} X B^{\mu}-i g_{s}G^{\mu})\Psi_{r}\,$. Notice that we adopted here a different normalization of the mixing terms with respect to Eq.~\eqref{generalMixing}. In Eq.~\eqref{2sitestop} we have neglected derivative interactions:\footnote{Derivative interactions can have a strong impact on the collider phenomenology of top partners \cite{DeSimone:2012ul,MatsedonskyiToAppear} as well as on EWPT \cite{Grojean:2013vw}.} these do not contribute to the potential nor affect the Higgs couplings (as discussed in the previous section), as long as all the parameters of the Lagrangian are real. In what follows we take the composite masses $M_{1,4,9}$ and linear mixings $F^{L,R}_{r}$ real. 

Integrating out the heavy fermions in the Lagrangian \eqref{2sitestop}, one obtains
\vspace{-1mm}
\be\l{2siteseff2}
\bry{lll}
 \La_{\text{eff}}^{t}&=&\dst \ovl{b}_{L}\,\psl\,\Pi_{b_L}(p) b_{L}+\ovl{t}_{L}\,\psl\, \Pi_{t_L}(p) t_{L}+\ovl{t}_{R}\,\psl\, \Pi_{t_R}(p) t_{R}\vspace{2mm}\\
&&\dst +\ovl{t}_{L} t_{R}\,\Pi_{t_L t_R}(p)+\text{h.c.}\,,
\ery\vspace{-2mm}
\ee
where the momentum-dependent form factors are
\vspace{-1mm}
\be\l{FFidentification}
\bry{ll}
\dst \Pi_{b_{L}}\,=&\dst \Pi_{0}^{b_{L}}+ \frac{1}{2}c_{h}^{2}\Pi_{2}^{b_{L}}\,,\vspace{2mm}\\
\dst \Pi_{t_{L}}\,=&\dst  \Pi_{0}^{t_{L}}+ \frac{1}{4}(1+c_{h}^{2})\Pi_{2}^{t_{L}} + s_{h}^{2}c_{h}^{2}\Pi_{4}^{t_{L}}\,,\vspace{2mm}\\
\dst \Pi_{t_{R}}\,=&\dst  \Pi_{0}^{t_{R}}+ \left(\frac{4}{5}-\frac{3}{4}s_{h}^{2}\right)\Pi_{2}^{t_{R}}+\frac{1}{20}\left(4-5s_{h}^{2}\right)^{2}\Pi_{4}^{t_{R}}\,,\vspace{2mm}\\
\dst \Pi_{t_{L}t_{R}}\,=&\dst  \frac{3}{4\sqrt{5}} M_{1} s_{h}c_{h}+\frac{1}{2\sqrt{5}} M_{2} s_{h}c_{h} \left(4-5s_{h}^{2}\right)\,,
\ery
\ee
with $s_{h}=\sin h/f$, $c_{h}=\cos h/f$ and
\be\l{FFs}
\bry{ll}
\dst\Pi_{0}^{b_{L},t_{L},t_{R}}= 1+\f{|F^{L,R}_9|^{2}}{p^{2}+M_{9}^{2}}\,,\vspace{2mm}\\
\dst\Pi_{2}^{b_{L},t_{L},t_{R}}=2\f{|F^{L,R}_4|^{2}} {p^{2}+M_{4}^{2}}-2\f{|F^{L,R}_9|^{2}} {p^{2}+M_{9}^{2}}\,,&\vspace{2mm}\\
\dst\Pi_{4}^{t_{L,R}}=\frac{5}{4}\f{|F^{L,R}_1|^{2}}{p^{2}+M_{1}^{2}}-2\f{|F^{L,R}_4|^{2}}{p^{2}+M_{4}^{2}}+\frac{3}{4}\f{|F^{L,R}_9|^{2}}{p^{2}+M_{9}^{2}}\,, &\vspace{2mm}\\
\dst M_{1}=2\left( \frac{ F^{L\ast}_{4}F^{R}_4 M_{4}}{p^{2}+M_{4}^{2}}-\frac{F^{L\ast}_{9}F^{R}_9 M_{9}}{p^{2}+M_{9}^{2}} \right)\,,&\vspace{2mm}\\
\dst M_{2}=  \hspace{-1mm}\left( \frac{5 F^{L\ast}_{1}F^{R}_1 M_{1}}{4(p^{2}+M_{1}^{2})}-\frac{2 F^{L\ast}_{4}F^{R}_4 M_{4}}{p^{2}+M_{4}^{2}}+\frac{3F^{L\ast}_{9}F^{R}_9 M_{9}}{4(p^{2}+M_{9}^{2})} \right).  &\vspace{2mm}\\
\ery
\ee
\normalsize
Integrating the path integral corresponding to the effective Lagrangian \eqref{2siteseff2} over the fermionic degrees of freedom, we can write the effective Coleman-Weinberg potential as
%
\small
\be\l{CWpotentialfermions}
V_{f}(h)=-2N_c\hspace{-1mm}\int\hspace{-1mm} \f{d^{4}p}{\(2\pi\)^{4}}\bigg[\log \Pi_{b_{L}} +\log\Big(p^{2}\Pi_{t_{L}}\Pi_{t_{R}}+\left|\Pi_{t_{L}t_{R}}\right|^{2}\Big)
\bigg],
\ee\normalsize
where $p$ is the Euclidean momentum 
and $N_{c}=3$ is the number of colors.
It is often convenient to expand the Higgs potential $V_{f}(h)$ in powers of $\epsilon\sim F/ M_{\Psi}\,$, where $F$ is a generic dimensionful linear mixing and $M_{\Psi}$ is some linear combination of the masses of the resonances. This expansion, however, breaks down for large compositeness, $\epsilon\sim 1\,$, which might be relevant for the top quark. Thus a more robust choice is to expand the potential in powers of $s^{2}_{h}$. Upon EWSB one has $s^{2}_{\<h\>}=\xi$ and since $\xi\ll1$ is required by EWPT, the expansion remains reliable even for $\epsilon\sim 1$. This expansion leads to
\vspace{-1mm}
\be\l{potetialexpxi}
V_{f}(h)\simeq a s_{h}^{2}+bs_{h}^{4}\,.
\ee
From the potential of Eq.~\eqref{potetialexpxi} we extract the values of the Higgs mass and {\scshape vev}
\vspace{-1mm}
\be\l{massvev}
\f{v^{2}}{f^{2}}=-\f{a}{2b}\,,\qquad m_{h}^{2}=\(\f{\partial^{2}V}{\partial h^{2}}\)_{\< h \>} = \f{8b}{f^{2}}\xi(1-\xi)\,.
\ee
It is important to note that the Higgs potential is quadratically divergent, unless the form-factors in \eq{FFs} fall off sufficiently fast at large Euclidean momenta. In what follows we propose two simple constructions where this is the case and the degree of divergence of the potential is reduced.

\vspace{-2mm}
\subsection{Two-Site Model}\label{sec:2site}
\vspace{-2mm}
%
%
One possibility to increase the calculability of the potential is to consider a two-site construction \cite{Panico:1359049} (see also Ref.~\cite{Redi:qi}). There, an unbroken $SO(5)$ global symmetry forces the relations $F^{L}_1=F^{L}_9=-F^{L}_4 \equiv F_{L}$ and $F^{R}_1=F^{R}_9=-F^{R}_4 \equiv - F_{R}\,$, and the quadratic divergences cancel. In this limit, the elementary/composite mixing terms in Eq.~\eqref{2sitestop} can be written as
\begin{equation}
-F_{L}\mathrm{Tr}[\overline{\mathcal{Q}}_{L}U^{T}\Psi_{R}U]-F_{R}\mathrm{Tr}[\overline{\Psi}_{L}U \mathcal{Q}_{R}U^{T}]+\mathrm{h.c.},
\end{equation}
where $\Psi$ is a complete $\mathbf{14}_{2/3}$ of composite fermions, see Eq.~\eqref{14decomp}. 
%
%
In the two-site construction, both $a$ and $b$ are \emph{logarithmically} divergent. We recall that we are expanding the potential in powers of $s_{h}^{2}$.\footnote{Notice that within this expansion one contribution to $b$ in Eq.~\eqref{potetialexpxi} is infrared divergent. We regulate this divergence with the top mass.} One more layer of resonances, corresponding to a three-site model, would be necessary to make the potential finite and therefore fully calculable. Instead, for illustrative purposes we regulate the potential by a cut-off $\Lambda\,$. This simple procedure allows us to estimate the value of the parameters in the potential and make qualitative predictions on the Higgs couplings and the corresponding spectrum of the resonances. The cut-off can be seen as roughly representing the mass scale of the third site (\emph{i.e.} of the second layer of resonances), but it is important to keep in mind that in our approach the logarithmic divergence also encodes finite terms, which can only be computed in a complete setup. For example, in a $5$-dimensional realization of the model we can expect $\Lambda \sim M_{KK}^{(2)}\sim 2M_{KK}^{(1)}$, where the KK modes are numbered with $1, 2,\ldots$. Since $M_{KK}^{(1)}$ is constrained from the $S$ parameter to be heavier than \mbox{$2\div3$ TeV} (the precise bound depending on the value of the $T$ parameter), we expect the cut-off scale $\Lambda$ to lie roughly between $5$ and $10$ TeV.

In order to perform a numerical study of the Higgs potential we fix $f=800$ GeV, $m_{t}\(\mu=1~\text{TeV}\)=152$ GeV and $v=246$ GeV and scan over the region of parameters\footnote{Notice that we assume $M_{9}>0$. Provided $M_{1},M_{4}$ can have both signs, the sign of $M_{9}$ can always be fixed without loss of generality.}
\be\l{parameters}
\bry{ccc}
M_{1},~M_{4}&\in&[-8,8] \text{ TeV}\,,\\
M_{9}&\in&[0,8] \text{ TeV}\,,\\
\Lambda&\in&[\max(|M_{1}|,|M_{4}|,M_{9}),10] \text{ TeV}\,,\\
F_{L}&\in&[0.1,6]f\,.\\
\ery
\ee
Notice that we do not scan over $F_{R}$, which is determined by the requirement that $m_{t}$ takes its experimental value. We require
\be\l{constraints}
\bry{ccc}
\xi&\in&[0.95,1.05]~v^{2}/f^{2}\,,\\
m_{h}&<& 160 \text{ GeV}\,,\\
\Lambda &>& \max(M_{\tilde{T}},M_{Q},M_{\psi})\,,\\
\min(M_{\tilde{T}},M_{X},M_{\psi})&>&500\,\text{GeV}\,.
\ery
\ee
The broad range of $m_{h}$ that we consider is motivated by the need of a sufficient statistics, but we expect that restricting the scan close to the measured value \mbox{$m_{h}\simeq 125$ GeV} would not qualitatively change our results. The masses that appear in Eq.~\eqref{constraints} are given by
\begin{eqnarray}
M_{\tilde{T}}=\sqrt{M_{1}^{2}+F_{R}^{2}}\,,\quad  M_{\psi} = |M_{9}|\,,\nonumber\\
M_{X}=|M_{4}|\,,\quad M_{Q}=\sqrt{M_{4}^{2}+F_{L}^{2}}\,.
\end{eqnarray}
Neglecting EWSB effects, $M_{\tilde{T}}$ is the physical mass of the $\mathbf{1}$, whereas $M_{\psi}$ is the mass of the degenerate $\mathbf{9}\,$, which contains $\psi\,$, an $SU(2)_{L}$ triplet with $Y=5/3$ whose top component $\psi_{8/3}$ has electric charge $8/3\,$ (see Table \ref{Table:fermions4}). On the other hand, the $\mathbf{4}$ is split into two $SU(2)_{L}$ doublets: $X$ with $Y=7/6$ and mass $M_{X}\,$, containing in particular $X_{5/3}\,$, a state with electric charge equal to $5/3\,$, and $Q$ with $Y=1/6$, which mixes with the elementary $q_{L}$ and thus has mass $M_{Q}\,$. As a preliminary estimate of the bounds from direct searches for vector-like quarks at the LHC, in our scan we require that all resonances are heavier than $500\,\GeV$, see the last line of Eq.~\eqref{constraints}. The actual LHC constraints obtained from 8 TeV data are however stronger: the mass of the $X$ doublet is bounded to $M_{X}>770\,\GeV$ by a dedicated CMS search for the $X_{5/3}$ \cite{CMS-PAS-B2G-12-012}, whereas a CMS search for the singlet $\tilde{T}$ gives the bound $M_{\tilde{T}}\gtrsim 700\,\GeV$ \cite{CMS-PAS-B2G-12-015}. The constraint on the $\psi_{8/3}$ and thus on the $\mathbf{9}$ is even stronger, $M_{\psi}>1\,\TeV$ \cite{Pappadopulo:2013wt,MatsedonskyiToAppear}.         
\begin{figure*}[ht!]
\begin{center}
\hspace{-2mm}\includegraphics[width=0.38\textwidth]{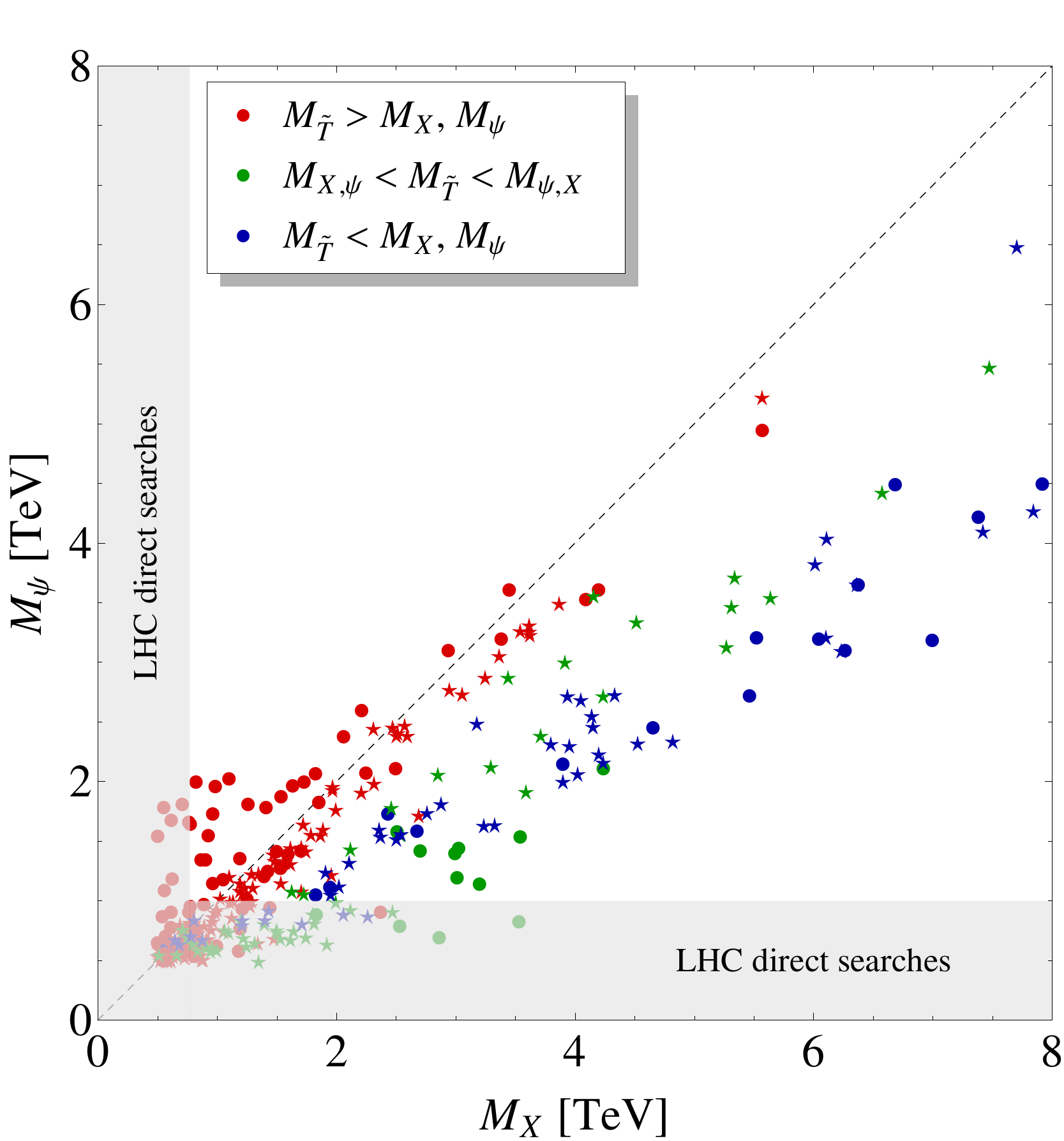}\hspace{18mm}
\centering\includegraphics[width=0.403\textwidth]{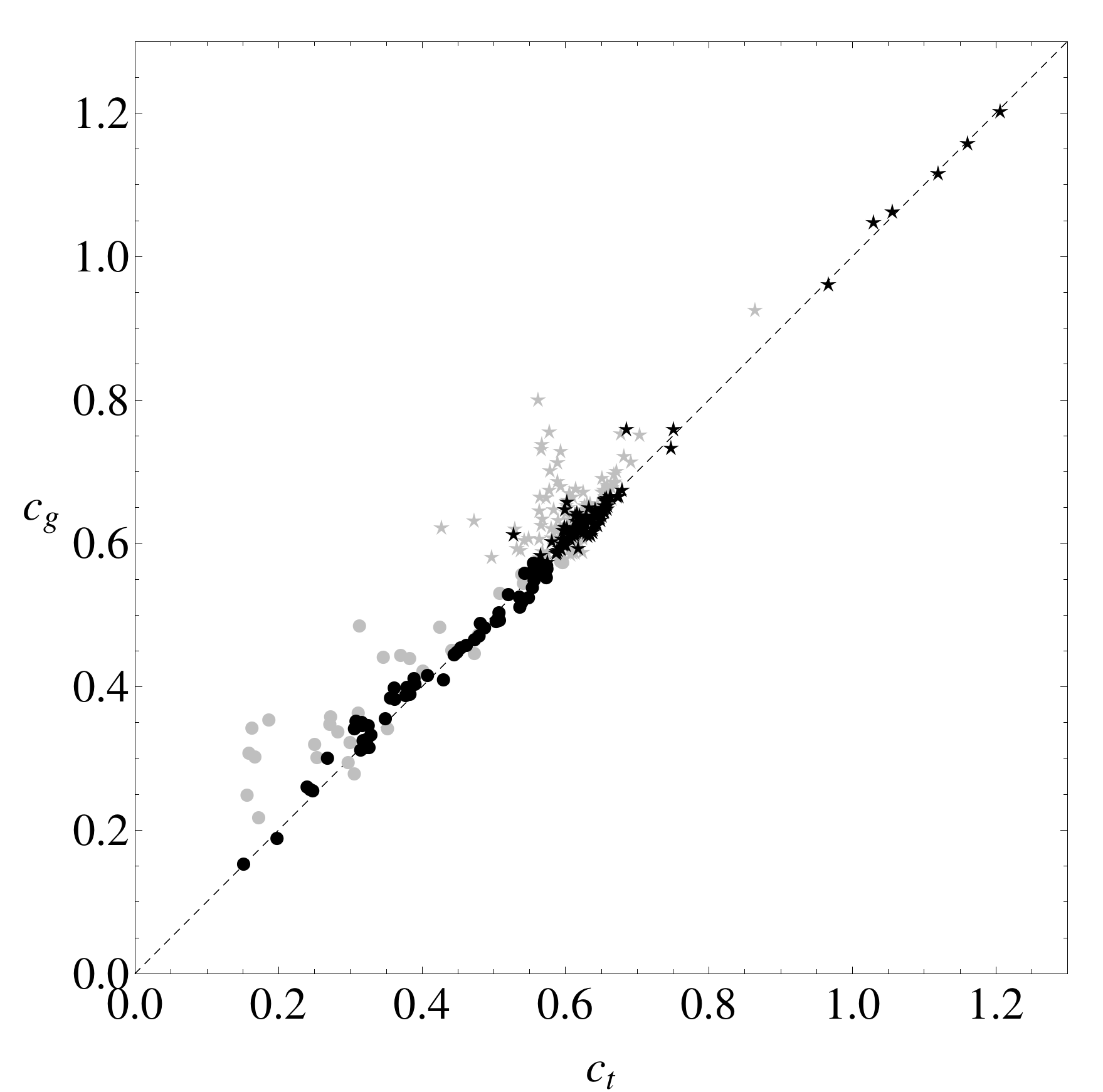}
\caption{\emph{Left panel:} distribution of the physical masses (neglecting EWSB corrections) of the $X$ doublet and of the degenerate $\mathbf{9}$. The current bounds from LHC direct searches are also displayed (there are no points ruled out only by the bound on the singlet $\tilde{T}$). The coloring of the points depends on the physical mass of the singlet. The points marked by a star are the ones for which the Higgs couplings $c_{g}$ and $c_{\gamma}$ are within the $95\%$ CL region of the fit to Higgs data, see Fig.~\ref{fig:fit}. \emph{Right panel:} Higgs coupling to gluons $c_{g}$ versus the Higgs coupling to the top quark $c_{t}$. Light gray points are excluded by LHC direct searches, while black points are currently allowed. The dashed line corresponds to the relation $c_{g}=c_{t}\,$, which holds for small mixings, $\epsilon^{2}\ll 1\,$. The meaning of the star shape for the points is the same as in the left panel.}
\label{fig:spectrum}
\end{center}\vspace{-5mm}
\end{figure*}

\begin{figure*}[th!]
\begin{center}
\includegraphics[width=0.45\textwidth]{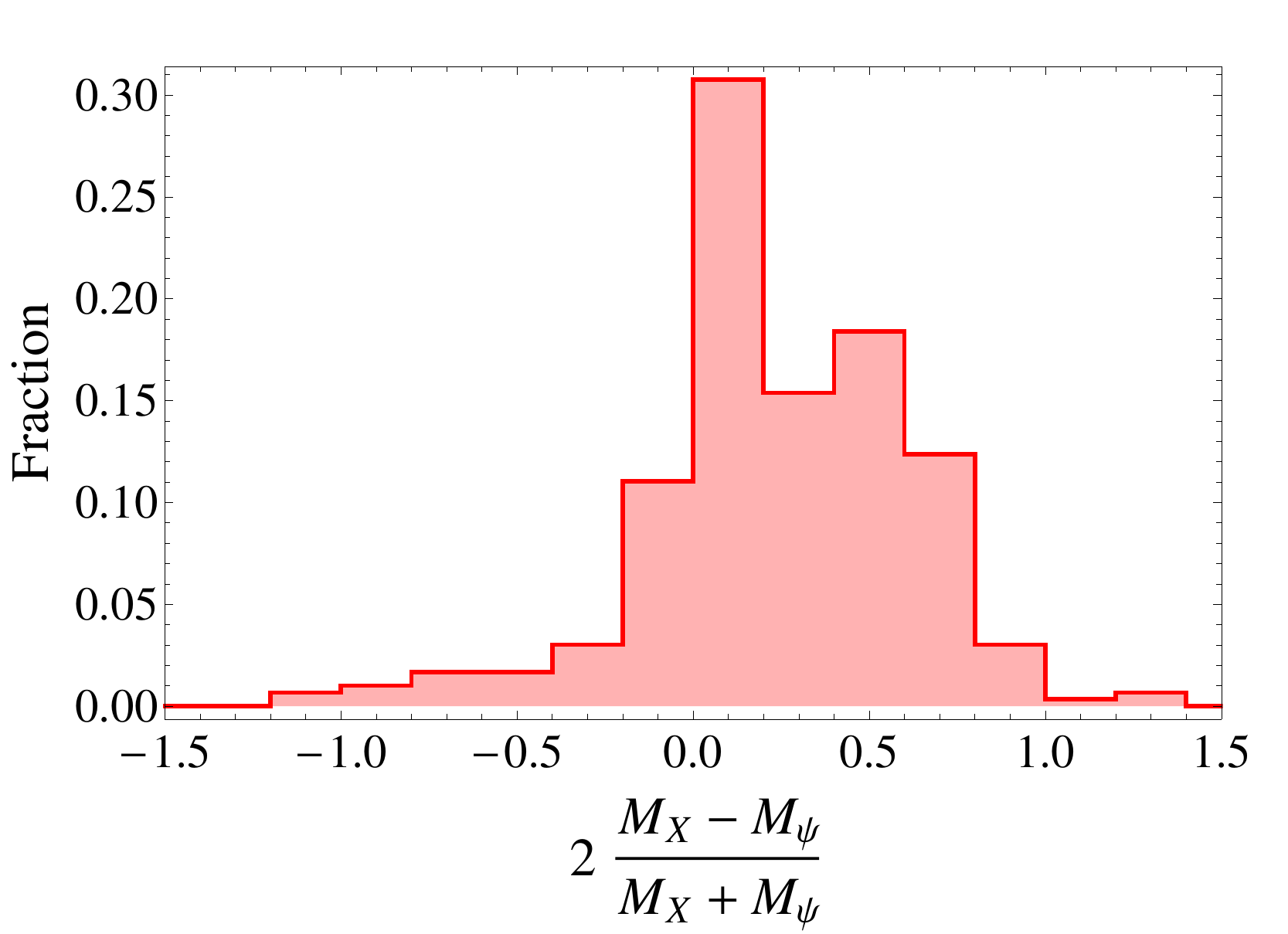}\hspace{10mm}
\centering\includegraphics[width=0.45\textwidth]{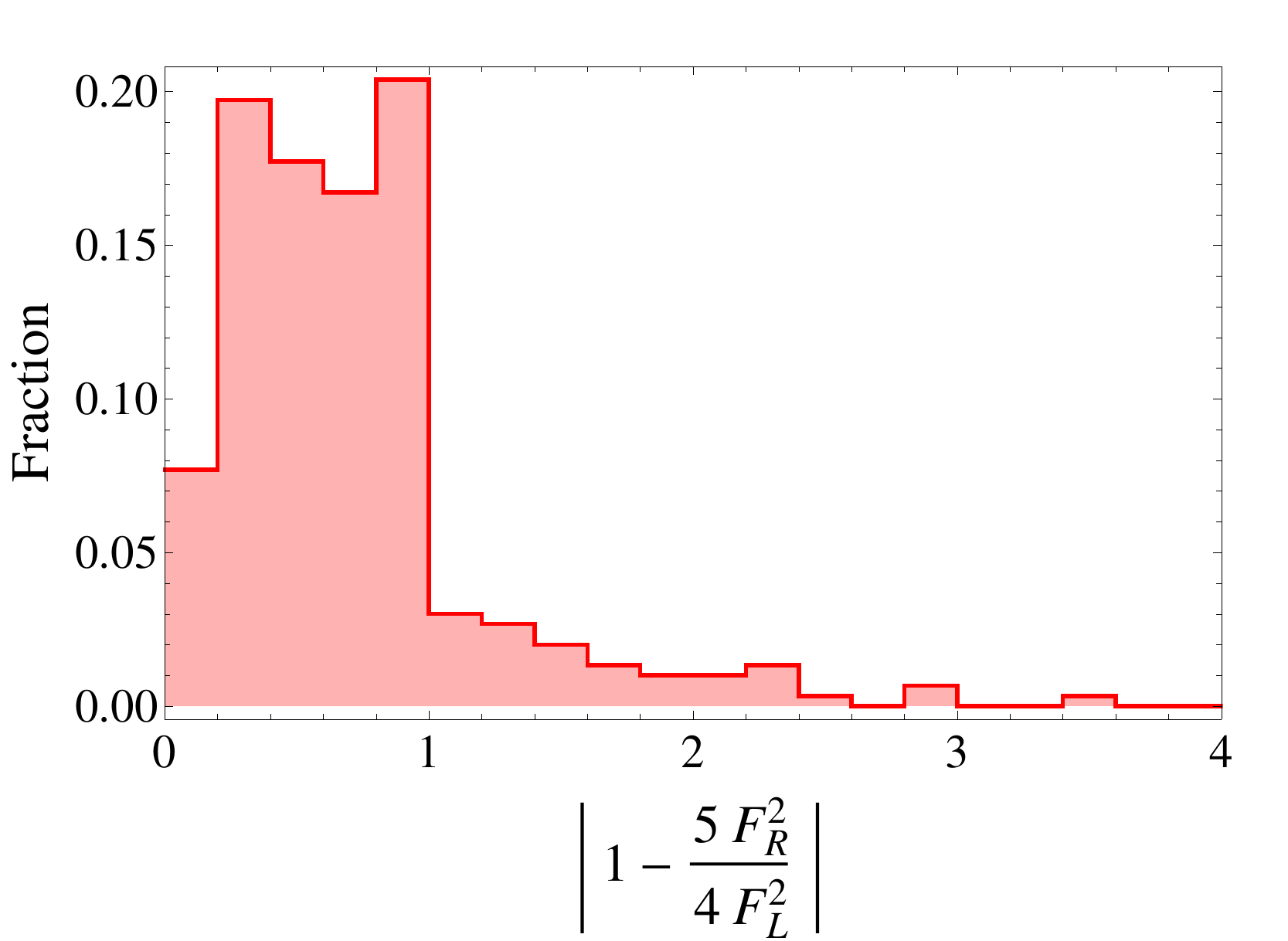}
\caption{Distributions quantifying the tuning between $M_{X}=|M_{4}|$ and $M_{\psi}=|M_{9}|$ (left panel), and between $|F_{L}|$ and $\sqrt{5}|F_{R}|/2$ (right panel), as obtained from the numerical scan.}
\label{fig:extratunings}
\end{center}\vspace{-5mm}
\end{figure*}

The spectrum of fermionic resonances as obtained from the scan is shown in Fig.~\ref{fig:spectrum}, together with the most up-to-date LHC constraints. The figure shows the values of $(M_{X},M_{\psi})$ for the points that satisfy all the requirements in Eq.~\eqref{constraints}, with a color code dependent on the mass of the singlet. The preferred spectrum is $M_{X}\sim M_{\psi} < M_{\tilde{T}}\,$, corresponding to the red points. Notice that in most of the viable parameter space the splitting between $M_{X}$ and $M_{\psi}$ is rather mild. This can be traced back to the expression the form factors in Eq.~\eqref{FFs} take in the two-site model: recalling that $M_{X,\psi}=|M_{4,9}|\,$, we see that for $M_{X}=M_{\psi}$ the form factors $\Pi_{2}^{b_{L},t_{L},t_{R}}$ exactly vanish. Thus for $M_{X}\sim M_{\psi}$ the overall size of the potential is suppressed, and a light Higgs is more likely obtained. In fact, in MCHM$_{14}$ two distinct invariants appear in the $O(\epsilon^{2})$ potential. This implies that only a tuning of order $\xi$ is necessary to obtain a realistic EWSB, as opposed for example to MCHM$_{5}$, where the tuning scales like $\epsilon^{2}\xi\,$. On the other hand, the potential in MCHM$_{14}$ is generically too large and yields a too heavy Higgs, unless some additional suppression mechanism is in play \cite{Panico:2012vr,Pomarol:2012vn,Pappadopulo:2013wt}. From our study of the two-site realization, we identify three main mechanisms that help in reducing the size of the Higgs mass. The first one is the already mentioned relation $M_{X}\sim M_{\psi}\,$. The second can be read from the expression of the Higgs mass at $\mathcal{O}(\epsilon^{2})$:
\begin{align}
m_{h}^{2}&\simeq \frac{2N_{c}}{\pi^{2}f^{2}}\xi\,\int dp\, p^{3} \left(\Pi_{4}^{t_{L}}-\frac{5}{4}\Pi_{4}^{t_{R}}\right) \nonumber \\
& \simeq \frac{2N_{c}}{\pi^{2}f^{2}}\xi\,\left(|F_{L}|^{2}-\frac{5}{4}|F_{R}|^{2}\right)\,M_{\Psi}^{2}\,,\label{mh154}
\end{align}
where $M_{\Psi}$ parameterizes the overall scale of the resonances. Thus for $|F_{L}|\sim \sqrt{5}\,|F_{R}|/2$ the leading contribution to the Higgs mass is suppressed. This relation is mildly satisfied in most of the viable parameter space. The last possibility is, of course, to lower the overall scale of the resonances $M_{\Psi}$. A combination of all three mechanisms is in play in our scan. The first two lead to extra tuning in addition to the one required for the Higgs {\scshape vev}. This extra tuning cannot be quantified from the scan, since we restrict ourselves to small regions around the realistic Higgs {\scshape vev} and mass. Nevertheless, as shown in Fig.~\ref{fig:extratunings}, the relations $M_{X}\sim M_{\psi}$ and $|F_{L}|\sim \sqrt{5}\,|F_{R}|/2$ are satisfied in a very mild sense, therefore we do not expect the consequent increase of the tuning to be dramatic.

The right panel of Fig.~\ref{fig:spectrum} shows the correlation between the Higgs coupling to gluons $c_{g}$ and the Higgs coupling to the top quark $c_{t}$. The former is computed from Eq.~\eqref{eq: hgg general} and reads at first order in $\xi$
\begin{equation} \label{eq:cg full}
c_{g}=1- \Delta_{g}^{(t)}\,\xi + \left(\frac{M_{4}^{2}}{M_{9}^{2}}-1\right)\sin^{2}\phi_{L}\, \xi\,, 
\end{equation}
\vspace{-1mm}
where 
\be
\Delta_{g}^{(t)} = \frac{11}{2}\,\frac{1-\frac{8}{11}\frac{M_{4}}{M_{1}}-\frac{3}{11}\frac{M_{4}}{M_{9}}}{1-\frac{M_{4}}{M_{1}}}
\ee
encodes the contribution of the top sector, whereas the last term is the contribution of the heavy $b$-like states. The angle $\phi_{L}\equiv \arctan (F_{L}/M_{4})$ measures the degree of compositeness of $q_{L}$. Notice that $\Delta_{g}^{(t)}$ only depends on ratios of the masses of the composite multiplets. The coupling of the Higgs to the top is obtained instead from Eq.~\eqref{eq: htt} with the identifications
\vspace{-1mm}
\be
m_{t}^{0}=-\Pi_{t_{L}t_{R}}(p=0)\,,\quad Z_{t_{L,R}}=\Pi_{t_{L,R}}(p=0)\,.
\ee
As discussed in the general analysis of Sec.~\ref{sec:general}, the $hgg$ and $ht\bar{t}$ couplings are tightly correlated, and significant deviations from the equality $c_{g}=c_{t}$ can occur only for large values of the mixing parameters $\epsilon$. This is clearly visible in the right panel of Fig.~\ref{fig:spectrum}: sizeable deviations from $c_{g}=c_{t}$ take place only for points that have already been excluded by direct searches at the LHC, displayed in light gray. For these points at least one of the masses $|M_{r}|$ is small, which typically implies that one of the mixings is large. For example, a small $|M_{4}|$ leads to large compositeness of $t_{L}$. In addition, we find that the corrections due to the wavefunction renormalization of the top are almost always negative, yielding $c_{t}\lesssim c_{g}\,$.   

\begin{figure}[t!]
\centering\includegraphics[width=0.4\textwidth]{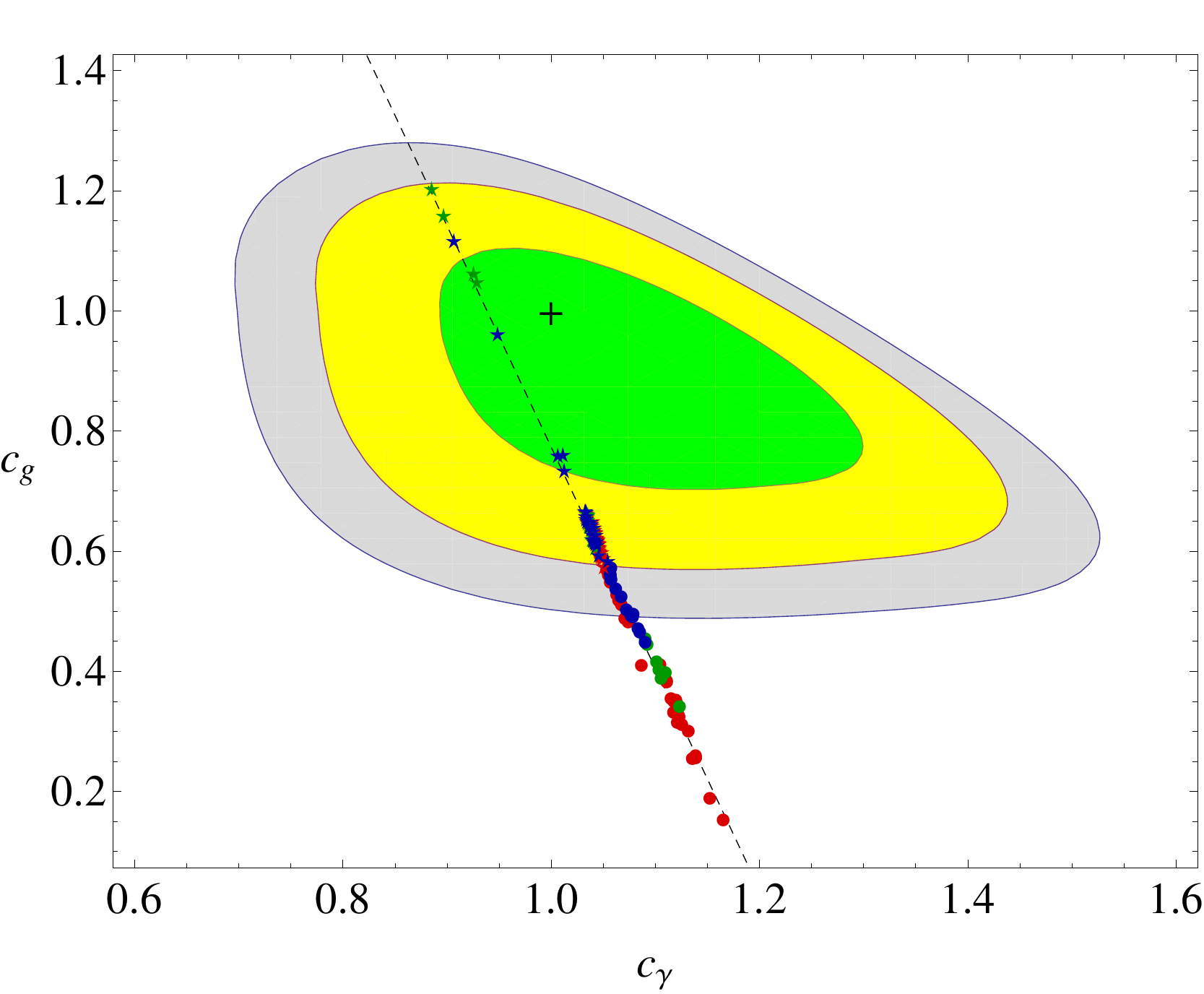}
\caption{Distribution of the couplings $(c_{\gamma},\,c_{g})$ as obtained from the scan, compared to the region preferred by a fit to current Higgs data. Only points not excluded by LHC searches for heavy vector-like quarks are displayed. The dashed line corresponds to the prediction of Eq.~\eqref{cgamma-cg analytical}. The green, yellow and gray regions correspond to the $68.27, 95$ and $99\%$ CL, respectively. As in Fig.~\ref{fig:spectrum}, the points marked by a star are those that fall within the $95\%$ CL region of the fit. Details on the fit can be found in Appendix~\ref{app:fit}.}
\label{fig:fit}\vspace{-3mm}
\end{figure}

In Fig.~\ref{fig:fit} we compare the Higgs couplings $(c_{\gamma},\,c_{g})$ obtained from the scan (considering only points not excluded by LHC direct searches) to the region preferred by a fit to current Higgs data. If only the contribution of fermions with electric charge $Q_{t}=2/3$ is considered, the points lie on the line
\begin{equation} \label{cgamma-cg analytical}
c_{g}=\left(1-\frac{7A_{V}(\tau_{W})}{4Q_{t}^{2}}\right)c_{\gamma}+\frac{7A_{V}(\tau_{W})}{4Q_{t}^{2}}\,c_{W}\,,
\end{equation}
where $\tau_{W}=m_{h}^{2}/(4m_{W}^{2})$ and $A_{V}(\tau_{W})\simeq 1.19$ parameterizes the $W$ loop \cite{Giardino:2013tu}, while $c_{W}=\sqrt{1-\xi}$ is the rescaling of the $hWW$ coupling in the MCHM. The loops of heavy $b$-like fermions generate only small deviations from this expectation.\footnote{Subleading corrections also arise due to the slightly different value of $\xi$ for each point.} Although a continuum of couplings is possible, Fig.~\ref{fig:fit} shows that there is a clear preference for $c_{g}\ll 1\,$, and as a consequence $c_{\gamma}>1\,$. Because we did not include the $b_{R}$ in our simple model, we cannot describe the $hb\bar{b}$ coupling, which plays an important role in the fit to data. Taking a model independent approach we remain agnostic on the sector that gives mass to the bottom quark, ignore the $b$ contribution to the $hgg$ and $h\gamma\gamma$ couplings and marginalize over the $hb\bar{b}$ coupling in the fit to Higgs data, see Appendix \ref{app:fit} for details.
\vspace{-3mm}
\subsection{Weinberg Sum Rules}
\vspace{-2mm}
Another possibility to obtain a finite Higgs potential, is to impose high-energy conditions on the form-factors $\Pi_{2}^{t_{L,R}}$ and $\Pi_{4}^{t_{L,R}}$ of \eq{FFs},
\vspace{-1mm}
\be\l{WSR}
\bry{ll}
\dst \lim_{p^2\to\infty} \Pi_{2,4}^{t_{L,R}}(p)\,&=\, 0\,,  \vspace{2mm}\\
\dst \lim_{p^2\to\infty} p^2\Pi_{2,4}^{t_{L,R}}(p)\,&=\, 0\,,\vspace{2mm}\\
\dst \lim_{p^2\to\infty} p^4\Pi_{2,4}^{t_{L,R}}(p)\,&=\, 0\,,
\ery
\ee
such that they fall-off rapidly at high momenta and the potential \eq{potetialexpxi} is convergent~\cite{Marzocca:2012tt,Pomarol:2012vn}. The QCD analog of the conditions \eqref{WSR} are known as Weinberg Sum Rules (WSR) \cite{1967PhRvL..18..507W}. Notice that in this approach we are expanding the potential in powers of $\epsilon^{2}$. Considering three resonance multiplets, as in \eq{2sitestop}, the conditions \eqref{WSR} cannot be satisfied simultaneously, but we can at least require that the Higgs mass be finite \cite{Pomarol:2012vn}. This can be done by imposing the conditions \eqref{WSR} only for $ \Pi_{4}^{t_{L,R}}$ which, as shown in \eq{mh154}, control the Higgs mass. Then, the WSR translate into relations between the couplings $F^{1,4,9}_{L,R}$ and the model is completely determined by the resonance masses and one combination of the couplings, which we chose to be $F^1_L$ (another combination, $F^1_R$ can be fixed by the top mass, $m_t\sim \langle s_hc_h\rangle F^1_L F^1_R/M_\Psi$). 

%
%
The general arguments given in section~\ref{sec:general} of course apply and in particular the coupling to gluons is independent of the mixing parameter $F^1_L$. We find
\vspace{-4mm}
\begin{equation}
c_{g}^{(t)}=1-\Bigg[4-\frac{M_1^3}{M_9^3}\frac{40M_4\left(1-\frac{M_4^2}{M_1^2}\right)-15M_9\left(1-\frac{ M_9^2}{M_1^2}\right)}{16 M_4 \left(1-\frac{M_4^2}{M_9^2}\right)-10 M_1 \left(1-\frac{M_1^2}{M_9^2}\right)}\Bigg]\xi\,,
\end{equation}
which reproduces the limits discussed below \eq{eq18}. Moreover, when any two resonances become degenerate, this expression simplifies to $c_{g}^{(t)}=1-3\xi/2\,$, so that \mbox{$c_{g}^{(t)}<1$} holds in most of the parameter space.

The coupling to tops $c_t$ differs from $c_g$ by terms proportional to $\epsilon_{L,R}^{2}\equiv(F^1_{L,R}/M_\Psi)^{2}$ which can in principle become sizable. Indeed,  while the product $F^1_L F^1_R$ is fixed by $m_t$, the Higgs mass is sensitive to another combination of the mixings,\footnote{Gluon partner contributions can modify this expression \cite{Barnard:2013tb}.}
\vspace{-4mm}
\begin{align}\label{mhWSR}
m^2_h\simeq&  \frac{5N_c}{4\pi^2f^2}\xi\left(|F^1_{L}|^2-\frac{5}{4} |F^1_{R}|^2\right)
\\
&\times\Big[M^2_{1}\log\left(\frac{M^2_{1}}{M^2_{9}}\right)
+\frac{M_{4}^2 (M^2_{9}-M^2_{1})}{M_{9}^2-M^2_{4}} \log\left(\frac{M^2_{9}}{M^2_{4}}\right)\Big]\, .\nonumber
\vspace{-5mm} \end{align}
This expression highlights how the Higgs mass can become small in this model; similarly to what discussed for the two-site construction, $m_h$ can be small if either \emph{i)} the overall scale of the resonances $M_\Psi$ is small or \emph{ii)} there is a tuning $|F_{L}^{1}|\sim \sqrt{5}\,|F_{R}^{1}|/2$ or \emph{iii)} a tuning between the masses $M_1\sim M_4$ or $M_1\sim M_9$. In the tuned cases \emph{ii)} and \emph{iii)} it is easy to see that the Higgs mass does not constrain the size of the $O(\epsilon_{L,R}^{2})$ corrections to $c_{t}$, and we can have a situation where $c_g$ and $c_t$ differ sizably. In the more natural case \emph{i)}, on the other hand, the $O(\epsilon_{L,R}^{2})$ corrections are typically small and $c_g\approx c_t$ holds.


\vspace{-4mm}
\section{Conclusions}\l{sec:conclusion}
\vspace{-3mm}
In composite Higgs models, the paradigm of partial compositeness implies that a number of colored fermionic resonances couple strongly to the Higgs sector. Moreover, some of these resonances need to be relatively light to naturally reproduce the observed Higgs mass. Thus one naively expects that these states contribute sizably to the radiative $hgg$ and $h\gamma\gamma$ couplings. However, it is well known that in some minimal models this is not the case and light fermionic resonances do not contribute to the $hgg$ and $h\gamma\gamma$ couplings, due to an exact cancellation between corrections to the $ht\bar{t}$ coupling and loops of resonances. Indeed the $hgg$ coupling is the leading term of the $ht\bar{t}$ one in an $\epsilon^{2}\ll1$ expansion.



In this paper we have shown that these are general features of the MCHM, following only from the Goldstone symmetry and from partial compositeness. Furthermore we found that under the assumption of $CP$ invariance the radiative Higgs couplings are insensitive to derivative interactions of the Higgs with resonances.\footnote{In models based on larger cosets the results of this paper would be modified, due to the presence of additional scalars that can mix with the Higgs \cite{Gripaios:2009pe,Mrazek:2011iu,Chala:2012te}.} Of particular interest for this generalization, are models where the top mass arises from more than one $SO(4)$ invariant. Such models, although disfavoured by the smallness of the Higgs mass, are particularly well-motivated by naturalness arguments.\footnote{More precisely, naturalness arguments prefer models in which more than one Left-Left (LL) or Right-Right (RR) invariant can be built, independently of the number of LR invariants. Nevertheless, the simplest models with more than one LR invariant, also feature more than one LL/RR invariants.} In this case, naively, the presence of  multiple operators can spoil the delicate cancellation that takes place in the simplest models. However, we found that this is not the case and the loop-induced Higgs couplings are insensitive to light fermionic resonances.

In the simplest models the $hgg$ coupling is reduced with respect to the SM value by a simple trigonometric factor ({\it e.g.}~$\cos(2\<h\>/f)/\cos(\<h\>/f)$ in the MCHM$_{5,10}$). On the contrary, in models with two or more invariants this coupling depends on the masses of the resonances and on their mixings with elementary fermions. In particular, it can become larger than the SM value and, for very special combinations of the parameters, it can differ from the SM value also in limit $v/f\to0$. Furthermore the coupling is insensitive to the overall scale of the resonances and only depends on ratios of their masses. Therefore one can imagine a situation where all the resonances are rather heavy and thus no signals show up in direct searches, but deviations are observed in the precision measurement of the Higgs couplings.


As an example, we have studied in detail a prototype model where both $q_{L}$ and $t_{R}$ are embedded in a {\bf 14} of $SO(5)$. We have built a two-site realization that enables the dominant part of the potential to be estimated, and used it to find a relation between the Higgs mass and {\scshape vev}, and the masses of the lightest resonances of the strong sector. In this simplified model, we have verified that $O(\epsilon)$ effects are small in the region of phenomenological interest. This confirms the tight connection between the $ht\bar{t}$ and the $hgg$ couplings. Moreover we find that these couplings are typically suppressed, leading also to a slight increase of the $h\gamma\gamma$ coupling. Similar qualitative conclusions have also been obtained by applying the WSR approach. The future direct measurement of the $ht\bar{t}$ coupling will provide a further test of these results. 

\vspace{-3mm}
\section*{Acknowledgments}
\vspace{-2mm}
We would like to thank L.~Da Rold, M.~Redi, J.~Serra and A.~Wulzer for useful discussions and A.~Azatov, R.~Contino, J.~Galloway and M.~Redi for comments about the manuscript. E.~S. has been supported in part by the European Commission under the ERC Advanced Grant 226371 {\it MassTeV}, the contract PITN-GA- 2009-237920 {\it UNILHC} and the ERC Advanced Grant 267985 {\it DaMeSyFla}. F.~R. acknowledges support from the Swiss National Science Foundation, under the Ambizione grant PZ00P2 136932 and thanks IFAE, Barcelona, for hospitality during completion of this work. The work of R.~T. was supported by the ERC Advanced Grant no. 267985 {\it DaMeSyFla} and by the Research Executive Agency (REA) of the European Union under the Grant Agreement number PITN-GA-2010-264564 {\it LHCPhenoNet}. F.~R., E.~S. and R.~T. thank the Galileo Galilei Institute for Theoretical Physics, Florence, for hospitality and the INFN for partial support during the completion of this work. We finally thank Heidi for computing resources and the grant SNF Sinergia n. CRSII2-141847.

\appendix
%
%
%
%
%
%
%
%
%
\vspace{-4mm}
\section{Notations} \l{app:notations}
\vspace{-2mm}
\subsection{Sigma model}
\vspace{-3mm}
The generators of the fundamental representation of $SO(5)$ read
\begin{align}
\dst T^{a\,L,R}_{IJ}\,=&\, -\frac{i}{2}\left[\frac{1}{2}\epsilon^{abc}(\delta_{I}^{\,b}\delta_{J}^{\,c}-\delta_{J}^{\,b}\delta_{I}^{\,c})\pm (\delta_{I}^{\,a}\delta_{J}^{\,4}-\delta_{J}^{\,a}\delta_{I}^{\,4})\right]\,, \nonumber \\
\dst T_{IJ}^{i}\,=&\, -\frac{i}{\sqrt{2}}(\delta_{I}^{\,i}\delta_{J}^{\,5}-\delta_{J}^{\,i}\delta_{I}^{\,5})\,, \l{generators5}
\end{align}
where $I,J = 1,\ldots,5$, $i = 1,\ldots,4$, \mbox{$a = 1,2,3$}. $T^{a\,L,R}$ are the generators of the unbroken \mbox{$SO(4)\sim SU(2)_{L}\times SU(2)_{R}$}, whereas $T^{i}$ are the generators of $SO(5)/SO(4)$. We will also use the equivalent notation $T^{a}\,$, $a=1,\ldots,6$ for the unbroken generators.
The Goldstone bosons appear through the matrix $U(\Pi)$ defined by
\be\label{GB}
U(\Pi)=\exp\left(i\frac{\sqrt{2} \Pi^{i} T^{i}}{f}\right)\,.
\ee
Notice that $U(\Pi)$ is an orthogonal matrix transforming as
\be\label{goldstones2}
U(\Pi)\to g\, U(\Pi) \, \hat h(g,\Pi)^{-1}, \qquad g \in SO(5),~\hat h \in SO(4)\,.
\ee
%
%
%
%
The quantities $d_\mu$ and $e_\mu$ are defined as the projections of the object $-U^T(A_\mu+i\partial_\mu)U$ onto the broken and unbroken generators respectively, such that $d_\mu$ transforms linearly as a $4$-plet, while $e_\mu$ shifts under the unbroken $SO(4)$. At lowest order in the chiral expansion, we have 
\begin{equation}
d^i_\mu=\frac{\sqrt{2}}{f}\,\nabla_\mu \Pi^{i} + \ldots\,,\quad  e^{a}_\mu = - g A_\mu^a + \ldots \,
\end{equation}
with $\nabla_\mu \Pi^{i}=\partial_{\mu}\Pi^{i}-i A_{\mu}^{a}(T^{a})^{i}_{\,\,j}\,\Pi^{j}\,$. $A_{\mu}^{a}$ contains the vector fields associated to the gauged generators $T_{L}^{a}$ and $T_{R}^{3}$ in the unbroken $SO(4)$. See for example Ref.~\cite{DeSimone:2012ul} for the complete expressions. At the two-derivative level the Goldstone Lagrangian reads
\begin{equation}\l{Lchi}
\mathcal{L}= \frac{f^{2}}{4} d_{\mu}^{\,i} d^{\,i\,\mu}\,.
\end{equation}
In the unitary gauge where $\Pi^{1}=\Pi^{2}=\Pi^{3}=0$ and $\Pi^{4}=h$, we have simply
%
\begin{align}
U \,=\,& \(\bry{ccc|cc}
& & \vspace{-3mm}& & \\
 & \mathbb{I}_{3} & & &  \\
  & & \vspace{-3mm}& &  \\ \hline
   & & & c_{h} & s_{h} \\
    & & & -s_{h} & c_{h} \ery\)\,,
\end{align}\\
where we defined $s_{h}\equiv \sin (h/f)$ and $c_{h} \equiv \cos(h/f)$. The two-derivative Lagrangian \eqref{Lchi} can now be written as
\begin{equation}
\mathcal{L}=\frac{1}{2}\partial_{\mu}h\partial^{\mu}h+\frac{g^{2}f^{2}}{4}s_{h}^{2}\left[W^{+}_{\mu}W^{-\,\mu}+\frac{1}{2\cos^{2}\theta_{w}} Z_{\mu}Z^{\mu}\right],
\end{equation}
which fixes, once we identify the $W$ mass,
\be
\xi\equiv \f{v^{2}}{f^{2}} = \sin^{2}\f{\<h\>}{f}\,.
\ee

\subsection{Fermion representations}
\begin{table*}[ht!]
\begin{center}
\begin{tabular}{c|cccc}
$\mathbf{1}_{2/3}$&$T_{3}^{L}$ &$T_{3}^{R}$ & $Y$ & $Q$\\ \hline
\blue{$\tilde{T}$}& \blue{$0$} & \blue{$0$} & \blue{$\f{2}{3}$} & \blue{$\f{2}{3}$} \\
\multicolumn{3}{c}{}\\
\multicolumn{3}{c}{}\\
\multicolumn{3}{c}{}\\
\multicolumn{3}{c}{}\\
\multicolumn{3}{c}{}\\
\multicolumn{3}{c}{}\\
\multicolumn{3}{c}{}\\
\multicolumn{3}{c}{}\\
\end{tabular}\hspace{1cm}
\begin{tabular}{c|cccc}
$\mathbf{4}_{2/3}$&$T_{3}^{L}$ &$T_{3}^{R}$ & $Y$ & $Q$ \\ \hline
\red{$T$}& \red{$+\f{1}{2}$} & \red{$-\f{1}{2}$} & \red{$\f{1}{6}$} & \red{$\f{2}{3}$}  \\
\red{$B$}& \red{$-\f{1}{2}$} & \red{$-\f{1}{2}$} & \red{$\f{1}{6}$} & \red{$-\f{1}{3}$}  \\
$X_{5/3}$& $+\f{1}{2}$ & $+\f{1}{2}$ & $\f{7}{6}$ & $\f{5}{3}$ \\
$X_{2/3}$& $-\f{1}{2}$ & $+\f{1}{2}$ & $\f{7}{6}$ & $\f{2}{3}$  \\
\multicolumn{3}{c}{}\\
\multicolumn{3}{c}{}\\
\multicolumn{3}{c}{}\\
\multicolumn{3}{c}{}\\
\multicolumn{3}{c}{}\\
\end{tabular}\hspace{1cm}
\begin{tabular}{c|cccc}
$\mathbf{6}_{2/3}$&$T_{3}^{L}$ &$T_{3}^{R}$ & $Y$ & $Q$ \\ \hline
$\chi_{1}$& $+1$ & $0$ & $\f{2}{3}$ & $\f{5}{3}$  \\
$T_{1}$& $0$ & $0$ & $\f{2}{3}$ & $\f{2}{3}$  \\
$B_{1}$& $-1$ & $0$ & $\f{2}{3}$ & $-\f{1}{3}$  \\
$\chi_{2}$& $0$ & $+1$ & $\f{5}{3}$ & $\f{5}{3}$  \\
\blue{$T_{2}$}& \blue{$0$} & \blue{$0$} & \blue{$\f{2}{3}$} & \blue{$\f{2}{3}$} \\
\fuchsia{$B_{2}$}& \fuchsia{$0$} & \fuchsia{$-1$} & \fuchsia{$-\f{1}{3}$} & \fuchsia{$-\f{1}{3}$} \\
\multicolumn{3}{c}{}\\
\multicolumn{3}{c}{}\\
\multicolumn{3}{c}{}\\
\end{tabular}\hspace{1cm}
\begin{tabular}{c|cccc}
$\mathbf{9}_{2/3}$&$T_{3}^{L}$ &$T_{3}^{R}$ & $Y$ & $Q$ \\ \hline
$\psi_{8/3}$& $+1$ & $+1$ & $\f{5}{3}$ & $\f{8}{3}$  \\
$\chi_{3}$& $0$ & $+1$ & $\f{5}{3}$ & $\f{5}{3}$  \\
$T_{3}$& $-1$ & $+1$ & $\f{5}{3}$ & $\f{2}{3}$  \\
$\chi_{4}$& $+1$ & $0$ & $\f{2}{3}$ & $\f{5}{3}$  \\
$T_{4}$& $0$ & $0$ & $\f{2}{3}$ & $\f{2}{3}$  \\
$B_{3}$& $-1$ & $0$ & $\f{2}{3}$ & $-\f{1}{3}$  \\
$T_{5}$& $+1$ & $-1$ & $-\f{1}{3}$ & $\f{2}{3}$  \\
$B_{4}$& $0$ & $-1$ & $-\f{1}{3}$ & $-\f{1}{3}$  \\
$\psi_{-4/3}$& $-1$ & $-1$ & $-\f{1}{3}$ & $-\f{4}{3}$  \\
\end{tabular}\hspace{1cm}
\end{center}\vspace{-2mm}
\caption{Electroweak quantum numbers of the fermion fields in the $\mathbf{1}_{2/3},\,\mathbf{4}_{2/3},\,\mathbf{6}_{2/3},\,\mathbf{9}_{2/3}$ representations of $SO(4)\times U(1)_{X}$. In \red{red}, \blue{blue} and \fuchsia{fuchsia} we indicate the states with the SM quantum numbers of the \red{$q_{L}$}, \blue{$t_{R}$} and \fuchsia{$b_{R}$}.}\label{Table:fermions4}\vspace{-1mm}
\end{table*}
We report here for convenience the decomposition of the $SO(5)$ representations used in this paper in terms of $SO(4)$ multiplets. We have
\begin{equation}
\hspace{-2mm}\Psi_{\mathbf{5}}=\(\bry{cccc} \vspace{-3mm}\\ \Psi_{4} \\ \vspace{-3mm} \\ \hline  \Psi_{1} \ery\)\,,~~ \Psi_{\mathbf{10}}=\(\bry{ccc|c}  &&\vspace{-3mm}& \\ &\Psi_{6}&&\Psi_{4}/\sqrt{2} \\ &\vspace{-3mm}&& \\ \hline&-\Psi_{4}^{T}/\sqrt{2}& &0 \ery\)
\end{equation}
and
\begin{equation} \label{14decomp}
\Psi_{\mathbf{14}} =\(\bry{ccc|c}  &&\vspace{-3mm}& \\ &\Psi_{9}- \Psi_{1} \mathbb{I}_{4}/(2\sqrt{5})&&\Psi_{4}/\sqrt{2} \\ &\vspace{-3mm}&& \\ \hline&-\Psi_{4}^{T}/\sqrt{2}& &2\Psi_{1}/\sqrt{5} \ery\)\,,
\end{equation}
where $\Psi_{r}$ are $SO(4)$ multiplets. For the case $X=2/3$, the singlet and $4$-plet can be written as
\begin{equation}\l{composite1-4}
\Psi_{1_{\frac{2}{3}}}=\tilde{T}\,,\qquad \Psi_{4_{\frac{2}{3}}}=\frac{1}{\sqrt{2}}\begin{pmatrix} iB-iX_{5/3} \\ B + X_{5/3} \\ iT - X_{2/3} \\ i X_{2/3} - T \end{pmatrix}\,,
\end{equation}
while for the antisymmetric tensor we have
\begin{equation}
\Psi_{6_{\frac{2}{3}}}=\f{1}{2}\begin{pmatrix}
0  & \mathcal{T}_{12}^{+}  & i(\mathcal{B}_{12}^{+}-\mathcal{X}_{12}^{+}) & \mathcal{B}_{12}^{-}+\mathcal{X}_{12}^{-}  \\
  & 0 & \mathcal{B}_{12}^{+}+\mathcal{X}_{12}^{+} & i(-\mathcal{B}_{12}^{-}+\mathcal{X}_{12}^{-})  \\
 &  & 0 & -i\mathcal{T}_{12}^{-}  \\
 & & & 0 \end{pmatrix}\,,
\vspace{1mm}
\end{equation}
with $\mathcal{T}_{12}^{\pm}\equiv T_{1}\pm T_{2}\,$, $\mathcal{B}_{12}^{\pm}\equiv (B_{1}\pm B_{2})/\sqrt{2}$ and \mbox{$\mathcal{\chi}_{12}^{\pm}\equiv (\chi_{1}\pm \chi_{2})/\sqrt{2}\,$}, and for the symmetric traceless tensor
%
\small
\begin{equation}
\Psi_{9_{\frac{2}{3}}}=\f{1}{2}\begin{pmatrix}
 \mathcal{P}^{+}-T_{4} & i\mathcal{P}_{-}  & \mathcal{B}_{34}^{+}+\mathcal{X}_{34}^{+} & \hspace{-1mm}-i\mathcal{B}_{34}^{-}+i\mathcal{X}_{34}^{-}  \\
  & \hspace{-4mm}-\mathcal{P}^{+}-T_{4} & -i\mathcal{B}_{34}^{+} + i\mathcal{X}_{34}^{+} & -\mathcal{B}_{34}^{-}-\mathcal{X}_{34}^{-}  \\
 &  & T_{4}-\mathcal{T}_{35}^{-} & i\mathcal{T}_{35}^{+}  \\
 & & & T_{4}+\mathcal{T}_{35}^{-} \end{pmatrix}\,,
\vspace{1mm}
\end{equation}
\normalsize
where $\mathcal{P}^{\pm}\equiv \psi_{8/3}\pm i \psi_{-4/3}\,$, $\mathcal{B}_{34}^{\pm}\equiv (B_{3}\pm i B_{4})/\sqrt{2}\,$, \mbox{$\mathcal{\chi}_{34}^{\pm}\equiv (\chi_{4}\pm i \chi_{3})/\sqrt{2}$} and $\mathcal{T}_{35}^{\pm}\equiv T_{3}\pm T_{5}\,$. The decomposition of the $SO(4)$ multiplets in terms of fermions with definite electroweak quantum numbers is given in Table~\ref{Table:fermions4}.
\vspace{-3mm}
\section{Details of the experimental fit}\label{app:fit}
\vspace{-2mm}

The best option to compare these models with experiments, as we do in Fig.~\ref{fig:fit}, is to present the data as extracted assuming  modified couplings between the Higgs and the SM states. In composite Higgs models the Higgs couplings to $V=W, Z$ are shifted by $\delta c_V\simeq -\xi/2$, which we fix in our analysis assuming $f=800\GeV$, and hence $\xi\approx 0.1$. Higgs couplings to bottom quarks, on the other hand can vary considerably in general models (they can have a parametric form similar to $h\bar t t$ couplings, approximately corresponding to  \eq{eq: hgg 5+5} or \eq{eq18}) but typically are smaller than one (in units of the SM coupling). For this reason we parametrize our theoretical ignorance by marginalizing over the $h\bar bb$ coupling in the region $c_b\in[0.5,1]$. Fig.~\ref{fig:fit} is then obtained by letting the effective $hgg$ and $h\gamma\gamma$ couplings vary (the $h\bar t t$ coupling, independently from its contribution to $hgg$ and $h\gamma\gamma$, is not yet measured with enough accuracy to change this picture considerably). 

The statistical analysis is performed using the latest signal strenght data given by the Tevatron experiments and by ATLAS and CMS at Moriond 2013 and soon after; a summary of the signal strengths in the individual channels can be found in Refs.~\cite{Giardino:2013tu,Falkowski:2013vg}. The signal strengths are assumed to follow a Gaussian distribution and we fit the data by minimizing a $\chi^2$ as described in detail in Ref.~\cite{Montull:2012ik}. We sum statistical and theoretical errors in quadrature and neglect possible correlation effects, which we find to be a reasonable approximation.

\bibliographystyle{mine}
\bibliography{bibliography}

\end{document}